\documentclass[10pt,journal,compsoc]{IEEEtran}
\ifCLASSOPTIONcompsoc
  \usepackage[nocompress]{cite}
\else
  \usepackage{cite}
\fi
\usepackage{amsmath}
\usepackage{graphicx}
\usepackage{algorithm2e}
\usepackage{amsthm}
\usepackage{amsfonts}
\usepackage{graphicx}
\usepackage{subfigure}
\usepackage{epsfig}
\usepackage{pifont}
\usepackage{setspace}
\usepackage{epsfig}
\usepackage{pifont}
\usepackage{setspace}
\usepackage{float}
\usepackage{stfloats}
\usepackage{stmaryrd}
\usepackage{perpage}
\MakeSorted{figure}
\MakeSorted{table}
\usepackage{caption}

\ifCLASSINFOpdf
\else
\fi
\begin{document}
%
\title{A Coarse-to-Fine Multiscale Mesh Representation And Its Applications}
%
%
%
%

\author{Hao-Chiang~Shao,~\IEEEmembership{Member,~IEEE}%
\IEEEcompsocitemizethanks{\IEEEcompsocthanksitem H.-C. Shao is with the Dept. Statistics and Information Science, Fu Jen Catholic University, Taiwan. \protect\\
E-mail: shao.haochiang@gmail.com, 142702@mail.fju.edu.tw
}
\thanks{This work was funded by the Academia Sinica Thematic Research Project: AS-102-TP-B09 ``Olfactory computation in the \textit{Drosophila} brain'', directed by Prof. C.-T. Chien and Prof. W.-L. Hwang.}
\thanks{\underline{Short Versions}: (1) Shao, Hwang, and Chen, ``A backward wavelet remesher for level of detail control and scalable coding'', in IEEE ICIP 2014, pp.5596-5600, \textit{DOI: 10.1109/ICIP.2014.7026132}; and, (2) Shao and Hwang, ``Deriving 3D shape properties by using backward wavelet remesher'', in IEEE GlobalSIP 2017, pp.191 - 195, \textit{DOI: 10.1109/GlobalSIP.2017.8308630}.}
}

%
%

\markboth{ArXiv Version}%
{Shao: A Coarse-to-Fine Mutliscale Mesh Representation}
%



\IEEEtitleabstractindextext{%
\begin{abstract}
We present a novel coarse-to-fine framework that derives a semi-regular multiscale mesh representation of an original input mesh via remeshing. Our approach differs from the conventional mesh wavelet transform strategy in two ways. First, based on a lazy wavelet framework, it can convert an input mesh into a multiresolution representation through a single remeshing procedure. By contrast, the conventional strategy requires two steps: remeshing and mesh wavelet transform. Second, the proposed method can conditionally convert input mesh models into ones sharing the same adjacency matrix, so it is able be invariant against the triangular tilings of the inputs. Our experiment results show that the proposed multiresolution representation method is efficient in various applications, such as 3D shape property analysis, mesh scalable coding and mesh morphing.
\end{abstract}

}

\maketitle

\IEEEdisplaynontitleabstractindextext

%
\IEEEpeerreviewmaketitle

\IEEEraisesectionheading{\section{Introduction}\label{sec:intro}}
\IEEEPARstart{T}{he} mesh surface model, sometimes called the wireframe model and considered as a representation of 3D image 
data, is widely-used in applications, such as animation, game design, facial recognition \cite{Blanz03}, scene reconstruction \cite{Gallip10}, medical modeling \cite{Zheng08}, and bio-image atlasing and visualization \cite{BRC_2010,vsp2012,Shao2014TBME}.  
Traditionally, fundamental mesh processing methods, e.g., remeshing, simplification, compression, and editing, have been developed as hierarchical decomposable operators, which allow users to make a change at a coarser resolution and refine it at a finer resolution. Accordingly, multiresolution and wavelet analysis have generated considerable interest in the field of mesh surface representation.

Lounsbery et al. \cite{Lou97} and Eck et al. \cite{Eck} extended Mallat's multiresolution wavelet analysis method \cite{Mallat} to the mesh surface space of an arbitrary topological type. The extension decomposes a complicated surface into a sequence of surfaces, each of which approximates the input surface at a lower resolution, and the coarsest approximation is generally referred to as the base mesh \cite{Lee98,wavelet4CG}. 
Mesh wavelet analysis requires that the input surface is a semi-regular mesh. However, in most cases, meshes are not semi-regular, irrespective of whether they are derived by 3D scanning devices or obtained from tiling 3D isosurfaces. Therefore, a remeshing step, called the remesher, which maps an arbitrary mesh to a semi-regular one, is required before wavelet analysis can be performed \cite{Guskov00,Kho01,wavelet4CG}. 
We call this conventional mesh surface analysis strategy \emph{forward wavelet approximation} because the remesher is applied first, followed by wavelet analysis in a fine-to-coarse manner. 
In general, the remeshing process begins from a base mesh, which is derived from simplification of the original input mesh. 
The isosurface defined by the original input mesh may therefore be approximated based on various remeshing results, each of which corresponds to a specific base mesh. 
Consequently, the wavelet decomposition derived by forward approaches cannot be invariant against triangular tiling, i.e., the topological information\footnote{In this paper, we say that two meshes have the same topological structure/information if they have the same number of vertices, edges, and faces and the same adjacency matrix.} of the isosurface defined by the original input mesh.

Instead of the forward approach, we can exploit a \textit{backward} process to synthesize the wavelet transform from a base mesh to the input mesh surface via subdivision, i.e., remeshing. Because the proposed backward approach operates in a coarse-to-fine manner, two major considerations need to be clarified in advance: (1) the way to select vertices for constructing the base mesh; and (2) where the new vertices should be interpolated at finer levels. 
Without imposing any constraint on the shape similarity of the base mesh, any decision about the first consideration would be correct. Users can manually select vertices from the input mesh to construct a base mesh; or they can derive a base mesh from the input by either a mesh simplification algorithm \cite{Hoppe96,Garl97} or a base mesh optimization algorithm \cite{marinov2004optimization,marinov2005automatic}. 
As for the second consideration, the piercing procedure \cite{Guskov00} provides a feasible solution.
Based on the piercing procedure, the problem can be solved by a subdivision scheme, which can derive displacement vectors to deliver newly interpolated vertices to where they are supposed to be on the isosurface defined by the original input mesh.

In this paper, we propose a \textit{backward wavelet remesher} (BWR) that uses the backward wavelet process to generate a semi-regular approximation of the original input mesh from a user-specified input base mesh. BWR has two properties. 
First, it derives the wavelet information by only implementing the remeshing process, whereas the forward approach requires two steps: remeshing and mesh wavelet analysis. 
Second, because BWR works in a coarse-to-fine manner, it guarantees that the multiresolution representation of two homomorphic meshes will share the same topological structure if the vertices' order and adjacency matrix of their base meshes are identical. Because of the second property, the multiresolution mesh representation derived by BWR is useful in a larger number of applications than compression and approximation.
When two sequences of multiresolution mesh approximations have the same topological structure, one mesh can be morphed into another at any level by simply using vertex interpolation. It is also straightforward to make a 3D structural comparison between the two sequences of the multiresolution meshes. Our experiment results demonstrate the efficacy of the two properties, and show that the PSNR-performance of the proposed approach is comparable to that of forward approaches for scalable coding.

The remainder of this paper is organized as follows. We review related work In Section \ref{review}; describe the backward wavelet framework in Section \ref{sec:FrmWk}; explain parameter determination in Section \ref{sec:parameters}; consider applications of the proposed approach in Section \ref{sec:app}; present our experiment results in Section \ref{sec:exp}; and discuss limitations of the approach and future work in Section \ref{sec:limit}. Section \ref{conclusion} contains our concluding remarks.

\section{Related Work} 
\label{review}

The concept of the mesh wavelet transform presented by Lounsbery et al. in 1997 \cite{Lou97} motivated subsequent multiresolution mesh representation studies. The authors focused on the relationship between the mesh wavelet transform and the inverse of the subdivision process. Their method processes a semi-regular mesh in a fine-to-coarse manner, and the analysis matrix required to decompose the mesh surface is the inverse of the synthesis matrix used in the lazy wavelet framework \cite{Lou97,wavelet4CG}:
\begin{equation}
\left[ \begin{array}{c} \mathbf{A}^{j}_{lazy}\\ 
        \mathbf{B}^{j}_{lazy}\end{array} \right]
 =  \left[ \begin{array}{c c} P^j_{lazy} & Q^j_{lazy}\end{array}\right]^{-1}
 =  \left[ \begin{array}{c c}  \mathbf{P}_n^j  & \mathbf{0} \\
  \mathbf{P}^j_m   & \mathbf{I} \end{array} \right]^{-1}  \mbox{.}
  \label{affinexx}
 \end{equation}
In Eq. (\ref{affinexx}), $\mathbf{A}^j_{lazy}$ and $\mathbf{B}^j_{lazy}$ are the analysis matrices; $P^j_{lazy}$ and $Q^j_{lazy}$ are the synthesis matrices; and $P^j_{lazy}$ is the subdivision matrix (described in Section \ref{sec:FrmWk}) that characterizes the subdivision scheme.

Based on \cite{Lou97}, Khodakovsky et al. designed the \textbf{Progressive geometry compression (PGC)} method \cite{Kho00}, which applies the mesh wavelet transform on the semi-regular remeshed approximation derived by \textbf{Mulitresolution adaptive parameterization of surfaces (MAPS)} \cite{Lee98} for progressive coding. 
Because the connectivity of a remeshed approximation is encoded with the subdivision method, Khodakovsky et al.  developed 
in \textbf{PGC} 
a hierarchical edge-tree structure (shown in Figure \ref{fig:coef_in_spiht}(b)) to link edges across different scales. With such a structure, it becomes possible to extend the conventional zero-tree coding method from 2D images to 3D meshes. 
Subsequently, Guskov et al. and Khodakovsky et al. designed \textbf{Normal meshes (NM)} approaches \cite{Guskov00, Kho01} to overcome the inconvenience of compressing three-entry coefficient vectors generated by the mesh wavelet transform. 
Based on the coarsest mesh derived by \textbf{MAPS} and a concomitant mapping function that records how vertices on the original input mesh are projected on all coarser resolution meshes and approximated via barycentric coordinate system, \textbf{NM} modifies the vertices' positions on the coarsest mesh to obtain its base mesh. From such a base mesh, hopefully, vertices on a finer level mesh can be represented as normal offsets of the previous coarser mesh. 
In other words, the \textbf{NM} algorithm generates a multiresolution mesh in which each level can be written as a normal offset from its coarser version. As a result, the wavelet information of such meshes can be represented as a scalar rather than a three-entry vector. 
There are other methods of multiresolution mesh representation. 
For example, Guskov et al. \cite{guskov1999multiresolution} and Botsch et al. \cite{Botsch04} developed algorithms based on Burt and Adelson's image pyramid decomposition \cite{BApyramid}; while Dong et al. \cite{dong2015multiscale} used a tight wavelet frame representation for mesh surface denoising. However, these methods are not closely related to mesh wavelet decomposition, so we will not discuss them further.

Conventional multiresolution analysis methods operate in a fine-to-coarse manner. 
Consequently, the multiscale representations derived by such approaches vary with the triangular tilings of the remeshed approximation of the original input. Hence, it is difficult for forward mesh wavelet analysis to derive a multiscale comparison of two homomorphic isosurfaces when their triangular tilings comprise different numbers of vertices, edges, and faces.  
For example, to characterize local shape variations of the human hippocampus via the spherical wavelet transform \cite{schroder1995spherical}, Nain et al. designed a very elaborate pre-processing routine for their source hippocampus mesh surface models \cite{Nain2007Multi}. To obtain a tiling-invariant multiresolution analysis result, Nain et al.'s approach uses conformal mapping to map the hippocampus models to a sphere and then remeshes the mapped models so that they share the same topological information. In the final step, the spherical wavelet transform is applied after registering the remeshing results. Accordingly, applications based on forward approaches have a bias toward multicale approximation issues, such as compression, progressive transmission and level-of-detail control of a single mesh model \cite{wavelet4CG}. 
They are less suitable for warping, registration and structural comparison of the multiscale properties of two or more isosurfaces.  

\begin{figure}[!h]
\begin{tabular} { p{200pt} }
\includegraphics[width=0.45\textwidth, keepaspectratio=true]{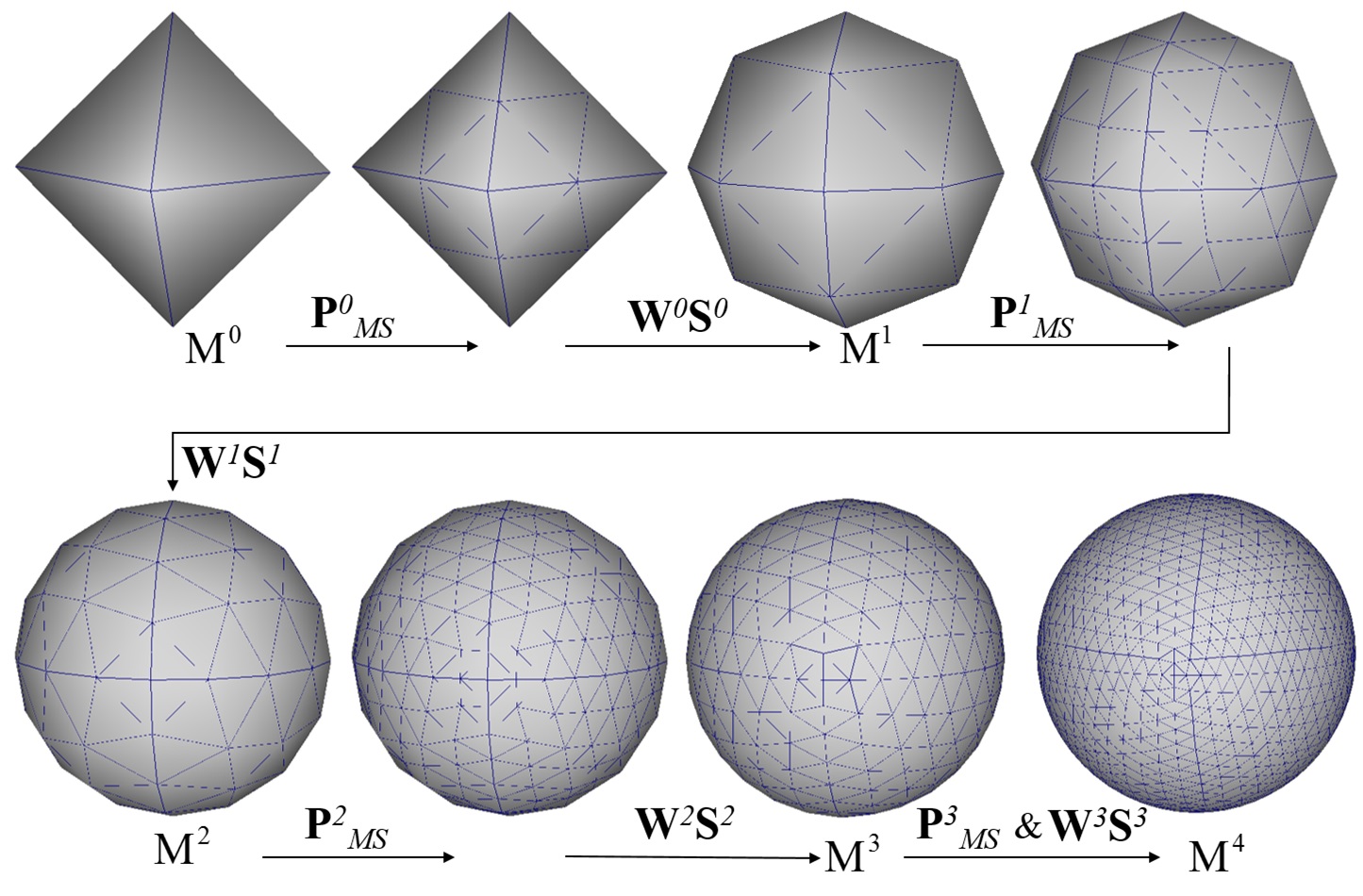}
\par \\[-0.7cm]
\end{tabular}
\caption{BWR representation of a sphere with an octahedron as the base mesh. $\textbf{P}^j_{MS}$
denotes the midpoint subdivisions, and ${\bf{W}}^j{\bf{S}}^j$ denotes the procedure of placing the midpoint vertices
on the surface of the original mesh.}
\label{fig:proposed_concept}
\end{figure}

To resolve the shortcomings of forward wavelet approaches, we posit that mesh wavelet analysis should be synthesized in a coarse-to-fine manner. The concept of the proposed method is illustrated in Figure  \ref{fig:proposed_concept}.
Briefly, detailed information about each resolution can be recorded by the subdivision coefficients $w$ and the  corresponding unit direction vectors $\vec{s}$. 
The latter should be adaptive to the local structure of the coarse resolution meshes.
The advantages of the proposed BWR are twofold. First, from the given base meshes with the same connectivity, the subdivision procedure can guarantee that the finer level approximation meshes will share the same adjacency matrix. Because of this \textit{tiling-invariant} property, BWR acts as a transformation procedure that can convert input meshes into a standard reference domain. Second, BWR derives the wavelet information directly from remeshing. By contrast, forward wavelet analysis is implemented in two steps: remeshing and mesh wavelet analysis. 
In addition, based on BWR, newly interpolated vertices on finer level meshes can be represented as a scalar denoting the length of the displacement vector, similar to that achieved by \textbf{NM}. The coding efficiency of BWR is therefore comparable to that of the \textbf{NM}-based method. 
In the next section, we review the frameworks of the conventional subdivision scheme and mesh wavelet transform, and then describe the proposed backward wavelet remesher (BWR) in detail.

\section{The Backward Wavelet Framework}
\label{sec:FrmWk}

Based on the lazy wavelet framework described in \cite{Lou97}, the $m^{th}$ vertex interpolated in the $j^{th}$ subdivision processes can be represented as the sum of its initial position $\tilde{\mathbf{q}}^j_m$ and a suitable displacement vector $\mathbf{w}^j_m$ denoting the detailed information; that is:
\begin{equation}
\label{eq:ACMTOG1997concept}
\mathbf{q}^j_m = \tilde{\mathbf{q}}^j_m + \mathbf{w}^j_m
\mbox{.}
\end{equation}
This equation is the basis for our backward wavelet remesher (BWR) design.
The detailed information about each resolution can be further represented as $\mathbf{W}^j \mathbf{S}^j$ (as shown in Figure \ref{fig:proposed_concept}) if the displacement vector $\mathbf{w}^j_m$ can be expressed as the product of a unit direction vector $\vec{s}^j_m$ and a scalar coefficient $w^j_m$, i.e., $\mathbf{w}^j_m = w^j_m \vec{s}^j_m$.

In the following subsections, we first review the matrix notations of the subdivision framework and conventional mesh wavelet transform and then describe the BWR method. This section provides the theoretical derivation to explain how we obtain a multiresolution mesh representation of an irregular input mesh by subdividing a user-specified base mesh.

\subsection{Subdivision and Its Matrix Notations} 
\label{subsec:subdvd}

Let $\mathrm{\textbf{V}}^{j}$ be the matrix of the coordinates of the vertices in the 
approximation subspace $\mathcal{M}^j$. The subdivision process determines the vertices in the finer approximation subspace $\mathcal{M}^{j+1}$ by applying a linear transform $\mathbf{P}^j$ on the vertices in $\mathcal{M}^j$ as follows:
\begin{equation}
\mathrm{\textbf{V}}^{j+1} = \mathrm{\textbf{P}}^{j} \,
\mathrm{\textbf{V}}^{j} = \left[
\begin{array}{c} \mathrm{\textbf{P}}^j_n \\ \mathrm{\textbf{P}}^j_m \end{array}
\right] \, \mathrm{\textbf{V}}^{j} \mbox{.} \label{subdivision}
\end{equation}
If the number of vertices in $\mathcal{M}^{j}$ is $n$ and that in $\mathcal{M}^{j+1}$ is $n+m$, then $\mathrm{\textbf{V}}^{j}$ and $\mathrm{\textbf{V}}^{j+1}$ are an $n \times 3$ matrix and an $(n+m) \times 3$ matrix respectively. Note that $m$ is the number of newly interpolated vertices in $\mathcal{M}^{j+1}$, and the $i$-th row of $\mathrm{\textbf{V}}^{j}$ denotes the coordinate vector of $v^{j}_i$, i.e., the $i$-th vertex in $\mathcal{M}^j$. 
Additionally, $\mathrm{\textbf{P}}^j$ is an $(n+m) \times n$ subdivision matrix that can be represented as two block submatrices, viz. $\mathrm{\textbf{P}}_n^j$ (of size $n \times n$) and $\mathrm{\textbf{P}}_m^j$ (of size $m \times n$). 
$\mathrm{\textbf{P}}^j_n$ is used to map the vertices from $\mathcal{M}^j$ to $\mathcal{M}^{j+1}$. When the subdivision is an interpolating scheme, $\mathrm{\textbf{P}}^j_n$ is an identity matrix; hence, all existing vertices in $\mathcal{M}^j$ will not be re-positioned in $\mathcal{M}^{j+1}$. 
In contrast, $\mathrm{\textbf{P}}^j_m$ attempts to insert new vertices based on $\mathcal{M}^j$ 's vertices coordinates and connectivities. 
For example, the $\mathrm{\textbf{P}}^j_{m}$ of the butterfly subdivision scheme is an $m \times n$ sparse matrix in which each row only contains eight non-zero entries with the values $\frac{1}{2}$, $2w$, or $-w$. The locations and values of non-zero entries in $\mathrm{\textbf{P}}^j_{m}$ are related to the vertices that are exploited to derive the coordinate of a new vertex, as expressed in Eqs. (\ref{butterfly})-(\ref{eq:vec_s1}) and illustrated in Figure \ref{fig:btrfly_fig}.

\subsection{Wavelet Transform Framework}
\label{subsec:WTonMesh}

In the wavelet transform framework, the coordinate matrix of the vertices in $\mathcal{M}^{j+1}$ is derived according to
\begin{equation}
\mathbf{V}^{j+1} = \mathbf{P}^j \mathbf{V}^j + \mathbf{Q}^j \mathbf{W}^j \mbox{.}
\label{wavelet}
\end{equation}
Here, $\mathbf{P}^j \mathbf{V}^j$ are obtained by applying a subdivision process on $\mathbf{V}^j$; and $\mathbf{Q}^j \mathbf{W}^j$ are the wavelet coordinate matrices \cite{Lou97} that contain the  ``detailed information'' of the newly interpolated vertices. 

Let $\mathbf{V}^{j+1} = \left[ \begin{array}{c} \mathbf{c}^{j+1}_n \\
\mathbf{c}^{j+1}_m
\end{array} \right]$, $\mathbf{V}^j = [\mathbf{c}_n^j]$, and $\mathbf{W}^j =
[\mathbf{d}_m^j]$. 
Then, Eq. (\ref{wavelet}) can be re-written as follows:
\begin{eqnarray}
\left[ \begin{array}{c} \mathbf{c}^{j+1}_n \\ \mathbf{c}^{j+1}_m
\end{array} \right] & = & \left[ \begin{array} {c c} \mathbf{P}^{j}
& \mathbf{Q}^{j} \end{array} \right] \, \left[ \begin{array} {c}
\mathbf{c}^j_n \\ \mathbf{d}^{j}_m \end{array} \right] \\
\label{twoscale}
 &=&
\left[ \begin{array} {c c} \mathbf{P}_n^{j} & \mathbf{Q}^j_n \\
                      \mathbf{P}_m^{j} & \mathbf{Q}^j_m
\end{array} \right] \,
\left[ \begin{array} {c} \mathbf{c}^j_n \\ \mathbf{d}^{j}_m
\end{array} \right] 
\mbox{,}
\end{eqnarray}
where $\mathbf{c}^{j+1}_n$ and $\mathbf{c}^j_n$ denote the coordinates of $n$ vertices in $\mathcal{M}^{j+1}$ and  $\mathcal{M}^j$ respectively; $\mathbf{c}^{j+1}_m$ are the newly inserted $m$ vertices in $\mathcal{M}^{j+1}$; and  $\mathbf{d}^j_m$ are the wavelet coordinate vectors. 
If we choose $\mathbf{Q}_n^j =
\mathbf{0}_{n\times m}$ and $\mathbf{Q}^j_m = \mathbf{I}_m$,
Eq. (\ref{twoscale}) can be expressed in a lazy wavelet framework and becomes
\begin{eqnarray}
\left[ \begin{array}{c} \mathbf{c}^{j+1}_n \\ \mathbf{c}^{j+1}_m
\end{array} \right] & = &
\left[ \begin{array} {c c} \mathbf{P}_n^j & \mathbf{0}_{n\times m} \\
                      \mathbf{P}_m^{j} & \mathbf{I}_m
\end{array} \right] \,
\left[ \begin{array} {c} \mathbf{c}^j_n \\ \mathbf{d}^{j}_m
\end{array} \right] \mbox{.}
\label{eq:eq07}
\end{eqnarray}
Consequently, 
\begin{eqnarray}
\left[ \begin{array}{c} \mathbf{c}^{j+1}_n \\
        \mathbf{c}^{j+1}_m \end{array} \right]
& = & \left[ \begin{array}{c}  \mathbf{P}_n^j \\
  \mathbf{P}^j_m \end{array} \right] \mathbf{c}^{j}_n +
  \left[\begin{array}{c} \mathbf{0}_{n \times 3} \\ \mathbf{d}_m^j \end{array} \right] \mbox{.}
  \label{eq:conceptbasedonlazy}
 \end{eqnarray}
Eq. (\ref{eq:conceptbasedonlazy}) shows that the mesh wavelet representation can be regarded as a subdivision process in which the vertices' coordinates in $\mathcal{M}^{j+1}$ are obtained from a linear combination of those in $\mathcal{M}^j$ plus the ``new" information in $\mathbf{d}_m^j$.

If we use an interpolating subdivision that lets $\mathbf{P}_n^j = \mathbf{I}_n$, we obtain
\begin{eqnarray}
\left[ \begin{array}{c} \mathbf{c}^{j+1}_n \\
        \mathbf{c}^{j+1}_m \end{array} \right]
& = & \left[ \begin{array}{c}  \mathbf{I}_n \\
  \mathbf{P}^j_m \end{array} \right] \mathbf{c}^{j}_n +
  \left[\begin{array}{c} \mathbf{0}_{n \times 3} \\ \mathbf{d}_m^j \end{array} \right]. 
\label{eq:bwr_concept}
 \end{eqnarray}
Eq. (\ref{eq:bwr_concept}) indicates that 
(1) the positions of the vertices in $\mathcal{M}^{j}$ are unchanged in $\mathcal{M}^{j+1}$ after the subdivision process because $\mathbf{c}_n^{j+1} = \mathbf{c}_n^j$; 
and (2) the positions of the newly inserted vertices in $\mathcal{M}^{j+1}$ are obtained by the subdivision process $\mathbf{P}_m^j\mathbf{c}_m^j$ and corresponding translation offset $\mathbf{d}_m^j$.
As the coordinates of the interpolated vertices are determined by the matrices $\mathbf{P}_m^j$ and $\mathbf{d}_m^j$, Eq. (\ref{eq:bwr_concept}) leads to the proposed backward wavelet remesher for multiresolution mesh representation.

\subsection{BWR Framework}
\label{subsec:bwr_frame}

Eq. (\ref{eq:bwr_concept}) implies that if a subdivision procedure can send the newly interpolated vertices $c_m^{j+1}$ to where they are supposed to be located, $c_m^{j+1}$ can be positioned on the target mesh surface without the detailed information $\mathbf{d}^{j}$.
That is, by letting $\mathrm{d}_m^{j}=0$, Eq. (\ref{eq:bwr_concept}) becomes
\begin{eqnarray}
\left[ \begin{array}{c} \mathbf{c}^{j+1}_n \\
        \mathbf{c}^{j+1}_m \end{array} \right]
& = & \left[ \begin{array}{c}  \mathbf{I}_n \\
  \mathbf{P}^j_m \end{array} \right] \mathbf{c}^{j}_n  
 =  \left[ \begin{array}{c}  \mathbf{c}^{j}_n \\
  \mathbf{P}^j_m \mathbf{c}^{j}_n \end{array} \right] \mbox{.} 
\label{eq:eq10}
 \end{eqnarray}
Note that, as mentioned in \cite{wavelet4CG}, there is no limitation on the subdivision method in conventional multiresolution mesh representation approaches. Our only restriction here is the presumption that $\mathbf{Q}_n^j = \mathbf{0}_{n\times m}$ and $\mathbf{Q}^j_m = \mathbf{I}_m$ (that is, the subdivision method itself must be an interpolating scheme), and the form of $\mathbf{P}^j_m$ does not matter. Therefore, the goal of BWR is to develop a subdivision procedure that can ensure (1) $\mathbf{c}_m^{j+1}$ are on the isosurface defined by the original input mesh $\mathcal{M}_{ref}$; and (2)  $\mathbf{c}_m^{j+1}$ will not fold patches and distort the geometry. 

\begin{figure}
\centerline{\includegraphics[width=7cm,height=5cm,keepaspectratio=true]{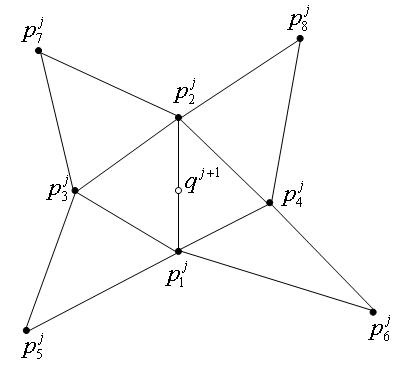}}
\caption{The eight neighboring vertices for butterfly subdivision interpolation of the vertex $q^{j+1}$.} \label{fig:btrfly_fig}
\end{figure}

Although the subdivision strategy required by BWR is inexplicit at this point, we should first check what will happen if the butterfly subdivision method \cite{Dyn90} is adopted. 
Let $\vec{q}_l^{j+1}$ be the initial coordinate of the $l$-th interpolated vertex $p_l^{j+1}$ in $\mathcal{M}^{j+1}$.
Then, the lower part of Eq. (\ref{eq:bwr_concept}), i.e., $\mathbf{c}^{j+1}_m=\mathbf{P}^j_m \mathbf{c}^j_n+\mathbf{d}^j_m$, can be rewritten in the vertex-wise form as follows:
\begin{equation}
\vec{p}_l^{j+1} = \vec{q}_l^{j+1} + \vec{d}^j_l \mbox{.}
\end{equation}
As shown in Figure \ref{fig:btrfly_fig}, the butterfly subdivision scheme yields
\begin{equation}
\vec{q}_l^{j+1} = \frac{1}{2}(\vec{p}^{j}_{l,1}+\vec{p}^{j}_{l,2}) + 2w_l(\vec{p}^{j}_{l,3}+\vec{p}^{j}_{l,4}) 
-
w_l(\vec{p}^{j}_{l,5}+\vec{p}^{j}_{l,6}+\vec{p}^{j}_{l,7}+\vec{p}^{j}_{l,8}),
\label{butterfly}
\end{equation}
where $w_l$ is a scalar parameter; and $\vec{p}_{l,1}^j \cdots \vec{p}_{l,8}^j \in \{\mathbf{c}_n^j\}$ are the eight points in the neighborhood of $p^{j+1}_l$ for the butterfly weighting average kernel.
Eq. (\ref{butterfly}) can then be re-written as
\begin{equation}
\vec{q}_l^{j+1} = \frac{1}{2}(\vec{p}^{j}_{l,1}+ \vec{p}^{j}_{l,2})
+ w_l^j \vec{s}_l^j, \label{butterfly1}
\end{equation}
where \begin{equation}\vec{s}_l = 2
(\vec{p}^{j}_{l,3}+\vec{p}^{j}_{l,4}) -
(\vec{p}^{j}_{l,5}+\vec{p}^{j}_{l,6}+\vec{p}^{j}_{l,7}+\vec{p}^{j}_{l,8}) \mbox{.}
\label{eq:vec_s1}\end{equation}
Therefore, we have
\begin{eqnarray}
\vec{p}_l^{j+1} & = & \vec{q}_l^{j+1} + \vec{d}^j_1 \\
& = &  \frac{1}{2}(\vec{p}^{j}_{l,1}+\vec{p}^{j}_{l,2}) + w_l
\vec{s}_l + \vec{d}^j_l \mbox{.}
\label{incident2}
\end{eqnarray}
The above equation states that the coordinate of the vertex $p_l^{j+1}$ is rooted at the mid-point of $\vec{p}^{j}_{l,1}$ and $\vec{p}^{j}_{l,2}$ and stretched in the direction $w_l \vec{s}_l + \vec{d}_l^j$.

Rewriting Eq. (\ref{eq:eq10}) according to Eq. (\ref{incident2}), 
the coordinates of the vertices in $\mathcal{M}^{j+1}$ can be represented as the following matrix:
\begin{eqnarray}
\left[ \begin{array}{c} \mathbf{c}^{j+1}_n \\
        \mathbf{c}^{j+1}_m \end{array} \right]
%
%
& = & \mathbf{P}^j_{mid} \mathbf{c}^{j}_n +
  \left[\begin{array}{c} \mathbf{0}_{n \times 3} \\ \mathbf{W}^j\mathbf{S}^j + \mathbf{d}^j \end{array}\right] \mbox{.}
\label{perturbation1}
 \end{eqnarray}
In Eq. (\ref{perturbation1}), $\mathbf{P}^j_{mid}$ is the mid-point subdivision kernel \cite{Book_subdvd}, defined as
\begin{equation}
\left[
\begin{array}{c }
\mathbf{I}_n^j \\ 
\mathbf{P}^j_{\frac{1}{2}} \\ \end{array} \right] \mbox{,}
\label{eq:interp_kernel}
\end{equation}
where $\mathbf{I}_n^j$ is an $n \times n$ identity matrix, and $\mathbf{P}^j_{\frac{1}{2}}$ is an $m \times n$ sparse matrix with two nonzero entries $\frac{1}{2}$ and $\frac{1}{2}$ in each row. 
$\mathbf{W}^j$ is an $m \times m$ diagonal matrix whose $l$-th diagonal element is $w_l$, 
and $\mathbf{S}^j$ is an $m \times 3$ matrix whose $l$-th row is the direction vector, $\vec{s}_l$, corresponding to $w_l$.
By letting $\mathbf{d}^j=0$, Eq.(\ref{perturbation1}) can be represented as a two-scale relationship of the wavelet transform as follows:
\begin{eqnarray}
\left[ \begin{array}{c} \mathbf{c}^{j+1}_n \\
        \mathbf{c}^{j+1}_m \end{array} \right]
& = & \mathbf{P}^j_{mid} \mathbf{c}^{j}_n +
  \left[\begin{array}{c} \mathbf{0}_{n \times 3} \\ \mathbf{W}^j\mathbf{S}^j\end{array} \right] \mbox{.} 
\label{eq:eq19}
 \end{eqnarray}
The matrix
\begin{equation}
  \left[\begin{array}{c} \mathbf{0}_{n \times 3} \\ \mathbf{W}^j\mathbf{S}^j\end{array} \right]
 \end{equation}
can be interpreted as a wavelet coordinate matrix. Eq. (\ref{eq:eq19}) shows a one-step, coarse-to-fine procedure for BWR. The locations of the vertices in ${\cal M}^j$ are kept the same in the finer subspace, a strategy that follows the interpolating subdivision scheme.
Meanwhile, the newly interpolated vertices are derived from the vertices in ${\cal M}^j$, modified with the mid-point subdivision matrix ${\bf{P}}_{mid}^j$ and the detailed information in ${\bf{W}}^j{\bf{S}}^j$. As shown in Figure  \ref{fig:proposed_concept}, the BWR construction is comprised of a sequence of midpoint subdivision procedures; and for each subdivision, the newly interpolated vertices are placed on the surface of the original mesh.

Eq. (\ref{eq:eq19}) implies that the problem of designing a proper $\mathbf{P}^j_m$, which is described in Eq. (\ref{eq:eq10}), can be converted to another problem of finding $\mathbf{S}^j$ and $\mathbf{W}^j$. 
In the next section, we describe how to derive these two matrices. Additionally, because $\mathbf{S}^j$ is evaluated according to the neighborhood of the newly interpolated vertices, the subdivision procedure of BWR is adaptive to local geometric information. Finally, we should explain that although we let  $\mathbf{d}^j=0$ in Eq.(\ref{eq:eq19}), this term should be preserved as per Eq. (\ref{perturbation1}) for possible future user-interactive extensions and robustness.

\begin{figure*}
\center
\begin{tabular} { p{0.30\textwidth}  p{0.30\textwidth}  p{0.30\textwidth}}
\includegraphics[height=4.1cm,keepaspectratio=true]{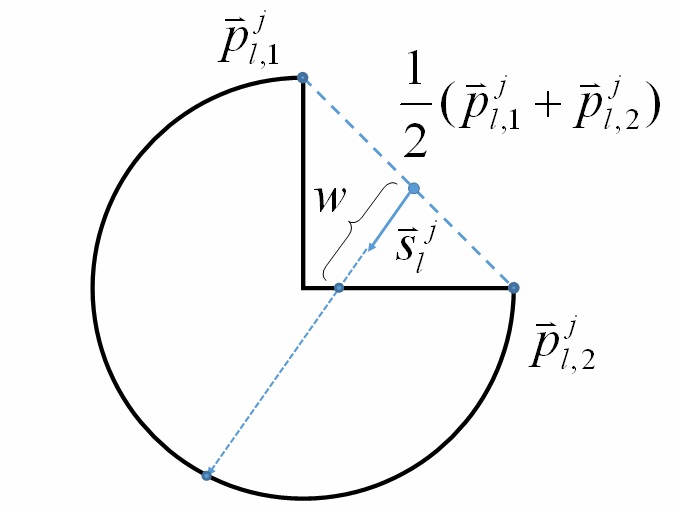}
\par &
\includegraphics[height=4.1cm,keepaspectratio=true]{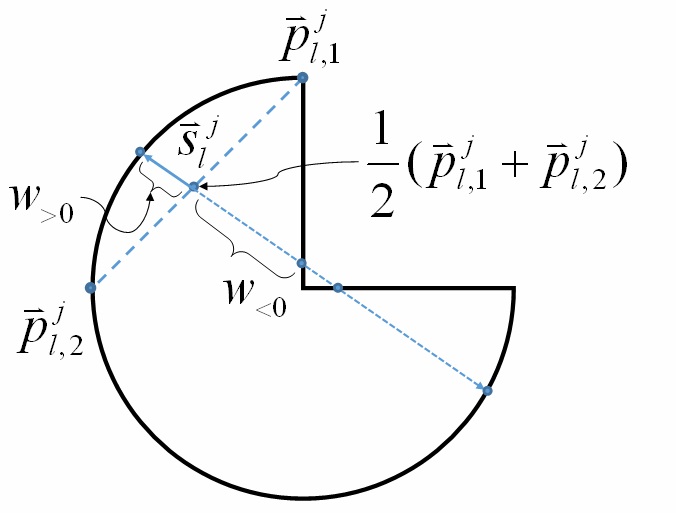}
\par &
\includegraphics[width=0.3\textwidth,height=4.1cm,keepaspectratio=true]{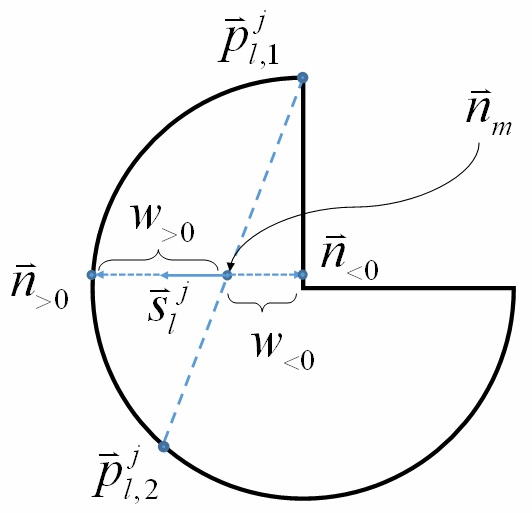}
\par \\[-0.6cm]
 \centerline{(a)} \par &
 \centerline{(b)} \par &
 \centerline{(c)} \par \\ [-0.6cm]
\end{tabular}
\caption{Determining the weight $w$ along the direction vector
$\vec{s}_l^j$.  The bold solid line is the contour of the original mesh; the
dashed line connecting $\vec{p}^j_{l,1}$ and $\vec{p}^j_{l,2}$ is a
portion of the coarser mesh $\mathcal{M}^j$; $\vec{s}^j_l$ is
located at the midpoint
$\frac{1}{2}(\vec{p}^j_{l,1}+\vec{p}^j_{l,2})$; and the new vertex
on $\mathcal{M}^{j+1}$ is at the intersection of
$\frac{1}{2}(\vec{p}^j_{l,1}+\vec{p}^j_{l,2})+ w \vec{s}_l^j$ and
the original mesh. If there are multiple intersections, the best
candidate is always one of the two nearest intersections, i.e., the
smallest $w>0$ and the largest $w<0$. 
(a) There are two intersections in the positive direction of $\vec{s}$ and the smallest $w>0$ is chosen. 
For (b) and (c), the intersections have one positive candidate (the smallest $w > 0$) and one negative candidate (the largest $w < 0$).
The value of $w$ is determined by checking whether $\vec{n}_{>0}$ or $\vec{n}_{<0}$ points in the direction of $\vec{n}_m$.
Here, $\vec{n}_m$ is the vertex normal of the midpoint
$\frac{1}{2}(\vec{p}^j_{l,1}+\vec{p}^j_{l,2})$; and $\vec{n}_{>0}$ and
$\vec{n}_{<0}$ are the normal vectors of the triangular patches associated with $w>0$ and $w<0$ respectively.
}
\label{fig:select_w}
\end{figure*}

\section{Parameter Determination} \label{sec:parameters}

BWR is comprised of three steps: selecting the vertices of the base mesh in the original mesh; determining the direction matrix $\mathbf{S}^j$; and deriving the weighting matrix $\mathbf{W}^{j}$ from the vertices in $\mathcal{M}^{j}$ and the original mesh.

\subsection{Selecting the Vertices of the Base Mesh in the Original Mesh}

The vertices of the base mesh must be chosen from the original mesh because BWR is based on the interpolating subdivision scheme. The base mesh can be derived by using a mesh simplification method \cite{Hoppe96,Garl97,Lee98}, a base mesh optimization algorithm  \cite{marinov2004optimization,marinov2005automatic}, or through user intervention. The base mesh and the original input mesh must have the same genus type.


\subsection{Direction Matrix $\mathbf{S}^j$}

To preserve the topological structure during synthesis, BWR requires that a newly interpolated vertex must intersect the original mesh in a way that preserves the neighborhood topology of the original mesh. This property is determined by the rows in the direction matrix $\textbf{S}^j$.

There are two scenarios in which a new vertex $p_l^{j+1}$ derived by the vertices $p_{l,1} \cdots p_{l,8}$ may not be located in the region enclosed by $\vec{p}_{l,1} \cdots \vec{p}_{l,4}$: 
1) when the direction vector $\vec{s}$ used to place the newly interpolated vertex lies on a flat or nearly flat area; 
and 2) when the neighborhood of the newly interpolated vertex covers sharp edges. 
In the first case,
Eq. (\ref{eq:vec_s1}) states that if $\vec{p}^j_{l,1}, \cdots, \vec{p}^j_{l,8}$ are coplanar or nearly coplanar, $\vec{s}_l^j$ will lie on (or almost lie on) the same plane. This implies that the interpolated vertex might be moved along some traverse direction of the plane of $\vec{p}^j_{l,1}, , \cdots, \vec{p}^j_{l,8}$. 
In the second case, the neighboring vertices on any side of the sharp edge may shift $\vec{s}_l^j$ to an improper direction, such as the traverse direction of the local patch, and cause patch folding or patch overlapping.

In both cases, the solution is to modify the direction vector
$\vec{s}_l^j$ in order to preserve the neighborhood topology in a
finer approximation subspace.
We determine if the interpolated vertex is located in a flat area or in a neighborhood of an edge; then the direction of the vertex is replaced by the normal vector of $\frac{1}{2}(\vec{p}_{l,1}+\vec{p}_{l,2})$ according to the method in \cite{wavelet4CG,meyer2003discrete}.

\subsection{Weighting Matrix $\mathbf{W}^j$}
\label{determineW}

The parameters in the diagonal weighting matrix $\mathbf{W}^j$ are chosen so that the newly interpolated vertices are incident on the original mesh. 
This is called the piercing procedure \cite{Guskov00}, and it can be achieved by setting the $(l,l)$-th
element of $\mathbf{W}^j$ as $w_l^j$ so that
\begin{eqnarray} \vec{c}_l^{j+1}
&=&  \frac{1}{2}(\vec{p}^{j}_{l,1}+\vec{p}^{j}_{l,2}) + w_l^j
\vec{s}_l^j 
\mbox{ 
(see Eq. (\ref{perturbation1}) and (\ref{eq:eq19}))
}
\label{incident}
\end{eqnarray}
and $\vec{c}_l^{j+1}$ belongs to the original mesh. The entry
$w_l^j$ is determined by extending the direction vector
$\vec{s}_l^j$ to intersect the original mesh. The steps of the procedure are as follows: 
1) find the hyperplane that includes a triangular patch in the original mesh; 
2) extend $\vec{s}_l^j$ to find the intersection point on the hyperplane; 
and 3) determine if the intersection point is inside the triangular patch. 
Let the
triangular patch in the original mesh be $\Delta v_{1} v_{2} v_{3}$.
The hyperplane that contains $\Delta v_{1} v_{2} v_{3}$ with a
normal vector $\vec{n}_{123}$ can be represented as follows:
\begin{equation}
\label{eq:plane1} 
\vec{n}_{123}^t (\vec{x}-\vec{v}_{1}) = 0,
\end{equation}
where $\vec{n}_{123}=(\vec{v}_{2} -\vec{v}_{1})\times(\vec{v}_{3} -
\vec{v}_{1})$, and ``\textit{t}'' denotes matrix transpose. The point where $\vec{s}_l^j$ intersects the
hyperplane is at $ \vec{c}_l^{j+1}=
\frac{1}{2}(\vec{p}^{j}_{l,1}+\vec{p}^{j}_{l,2}) + w_l^j
\vec{s}_l^j$. Here, $w_l^j$ can be derived by replacing $\vec{x}$ in
Eq. (\ref{eq:plane1}) with
$\frac{1}{2}(\vec{p}^{j}_{l,1}+\vec{p}^{j}_{l,2}) + w_l^j
\vec{s}_l^j$. If the intersecting point $\vec{c}^{j+1}_l$ is in the
triangle $\Delta v_{1} v_{2} v_{3}$, then $\vec{c}^{j+1}_l$ can be
represented by a barycentric coordinate system of $\vec{v}_1,
\vec{v}_2$, and $\vec{v}_3$; that is,
\begin{equation}
\vec{c}^{j+1}_l = \alpha \vec{v}_1 + \beta \vec{v}_2 + \gamma
\vec{v}_3,
\end{equation}
where $\alpha, \beta, \gamma \ge 0$ and $\alpha + \beta + \gamma =
1$.

The shape of the original mesh determines whether it can contain multiple valid intersections with triangles. 
As shown in Figure \ref{fig:select_w}, the best candidate is always one of the two nearest intersections: one from the smallest $w > 0$ and the other from the largest $w < 0$. 
The following procedure can be used to make the final decision. 
Let $\vec{n}_{>0}$ and $\vec{n}_{<0}$ be the normal vectors of two candidate triangular patches corresponding to the smallest $w > 0$ and the largest $w < 0$ respectively. 
In addition, let $\vec{n}_m$ be the normal vector at the midpoint $\frac{1}{2}(\vec{p}^{j}_{l,1}+\vec{p}^{j}_{l,2})$ (see Eq. (\ref{incident})). 
The triangular patch whose normal vector points in the direction of $\vec{n}_m$ is chosen to determine the value of $w_l^j$.

\section{Applications}
\label{sec:app}

In this section, we discuss two practicable applications of the proposed method, namely, scalable coding and morphing.

\subsection{Scalable Coding} \label{scalablecoding}

BWR can derive a multiresolution approximation of an original input mesh and represent the location of every newly interpolated vertex as a scalar. Consequently, in terms of storage efficiency, BWR is as efficient as the Normal Meshes (\textbf{NM}) method \footnote{In the \textbf{Normal Meshes} approach, as reported in \cite{Guskov00}, more than $90\%$ of the vertices can be represented as scalar and rest of them are still represented as vector.} \cite{Guskov00, Kho01} for scalable (progressive) coding applications. 

\begin{figure}
\includegraphics[width=8.5cm,height=8cm,keepaspectratio=true]{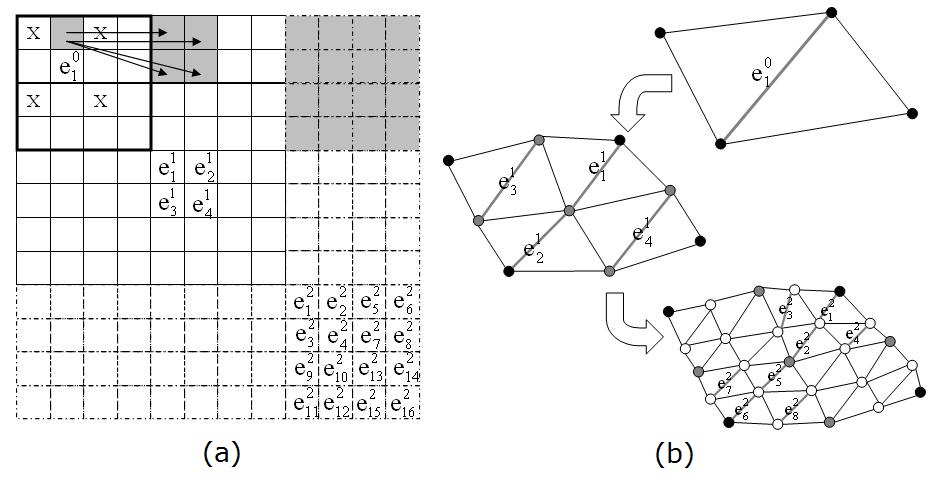}
\caption{
An example of mapping the vertices of a multi-resolution mesh to an image wavelet transform. The subdivision coefficient used to insert a new vertex in edge $e^{j}_i$ in (b) is placed in the pixel position, which is labeled $e^{j}_i$, in (a).
As shown in the mesh, the edge $e^0_1$ is subdivided into four edges: $e^1_1, e^1_2, e^1_3,$ and $e^1_4$; and the corresponding
coefficients are shown as children of $e^0_1$ in the image. In general, the children of $e^j_i$ are $e^{j+1}_{4i}, e^{j+1}_{4i-3},
e^{j+1}_{4i-2}$, and $e^{j+1}_{4i-1}$. The pixel labeled $\rm{x}$ is set at the mean of the three values in its underside, dextral, and diagonal positions.
} 
\label{fig:coef_in_spiht}
\end{figure}

Scalable coding can be regarded as a special case without any user input at the locations of the newly interpolated vertices in $\mathcal{M}^{j+1}$. This corresponds to setting $\mathbf{d}^j$ in Eq. (\ref{perturbation1}) as a zero matrix. 
Our mesh scalable coding framework 
regards the scalar matrices, $\mathbf{W}^j$'s, as a series of wavelet transform coefficients \cite{Lou97} and exploits Khodakovsky et al's edge-tree structure \cite{Kho00}. 
In addition, we map all the scalars (at different resolutions of the meshes) to the wavelet transform coefficients of an image. Then, we can use a common wavelet-based scalable codec for image compression to progressively transmit the vertices of our multiscale mesh representation. Progressive transmission can be performed by sending the coarsest mesh first, followed by the wavelet coefficients in a bitplane-by-bitplane manner for the refinement. 
Figure \ref{fig:coef_in_spiht} shows a simple example to demonstrate the parent-child relationship between a vertex at scale $j$ and its children at scale $j+1$. With the map of mesh vertices to image wavelet transforms, any wavelet-based scalable coding algorithm would be suitable for transmitting and encoding the information about the mesh vertices.

\subsection{Morphing}
\label{subsec:morphing}

Morphing is the process that gradually changes a source object through intermediate objects into a target object \cite{Lee1999SIGGRAPH}. 
In the conventional mesh morphing process, an essential step is to map both the source and the target meshes into the intermediate one. After deriving a function $\Pi_{i \rightarrow \alpha}$, which can map the isosurface defined by mesh-$i$ to that defined by the intermediate mesh, the morphing result  $\mathcal{M}_{\alpha}$ can be obtained according to Eq.(\ref{eq:morph_concept}):
\begin{equation}
\mathcal{M}_{\alpha}(v) = (1 - \alpha) \mathcal{M}_0 (\Pi^{-1}_{0 \rightarrow \alpha}(v))+ \alpha \mathcal{M}_1 (\Pi^{-1}_{1 \rightarrow \alpha}(v))
\mbox{,}
\label{eq:morph_concept}
\end{equation}
where $\mathcal{M}_0$ and $\mathcal{M}_1$ are the source and the target meshes respectively, $v$ denotes the vertices' coordinates, and $\Pi_{i \rightarrow \alpha}^{-1}$ is the inverse function of $\Pi_{i \rightarrow \alpha}$.

BWR can reduce the complexity of deriving the mapping function. Because BWR can remesh meshes of the same genus type into ones that share the same topological information, it is straightforward to construct a one-to-one mapping between vertices of two remeshed approximations.
Given two mesh models, $\mathcal{M}_A$ and $\mathcal{M}_B$, we can generate an intermediate mesh $\mathcal{M}_I(\alpha)$ between them based on a one-to-one mapping function $\mathbf{\Pi}$ by
\begin{equation}
v^I_i(\alpha) = \alpha \, v^A_i + (1-\alpha) \, v^B_j \, \mbox{ with } j=\Pi(i) \mbox{.}
\label{eq:eq24}
\end{equation}
Hence, with the aid of BWR, the conventional mesh morphing problem can be simplified to a ``mesh blending'' problem. For cases where metamorphism among multiple meshes needs to be derived simultaneously\footnote{For example, in order to produce a standard average atlas model for 3D image atlasing, one may have to register, warp and average multiple 3D models simultaneously, as described in \cite{Shao2014TBME}. Such an average can be considered as an equal-weighted metamorphism among all models.}, the blended intermediate meshes can be obtained by
\begin{equation}
\mathcal{M}^I_\alpha(v) = \sum_{i} \, \alpha_i \, \mathcal{M}^i(v) \mbox{,}
\end{equation}
with $ \sum_{i} \, \alpha_i = 1$ and $0 \leq \alpha_i \leq 1$.

\section{Experiment Results}
\label{sec:exp}

In this section, we demonstrate the multiresolution approximations of BWR on some meshes and compare their performance with that of other remeshers. We also compare the coding performance of several multi-resolution mesh representation methods and analyze the morphing results. In the following experiments the mesh-to-mesh distance was measured by a well-known method, namely, METRO \cite{METRO,METROwebsite}; and the bit-rate was measured by the bit-per-vertex (\textbf{bpv}).

\begin{figure*}[b]
\center
\begin{tabular}{ p{80pt}p{80pt} p{80pt} p{80pt}p{80pt}}
\centerline{\includegraphics[width=6.0cm,height=5.3cm,keepaspectratio=true]{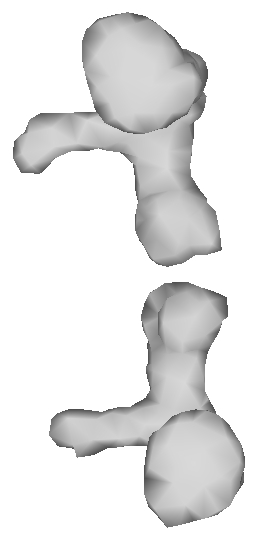}} \par &
\centerline{\includegraphics[width=6.0cm,height=5.3cm,keepaspectratio=true]{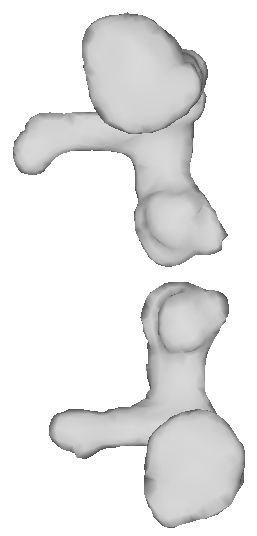}} \par &
\centerline{\includegraphics[width=6.0cm,height=5.3cm,keepaspectratio=true]{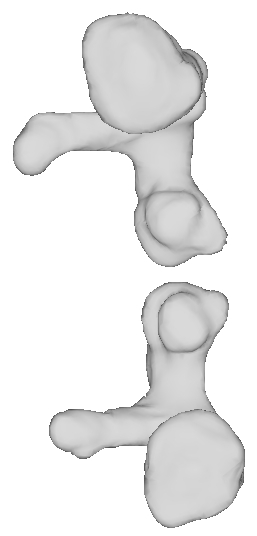}} \par &
\centerline{\includegraphics[width=6.0cm,height=5.3cm,keepaspectratio=true]{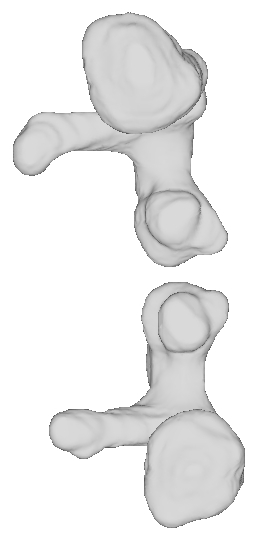}} \par &
\centerline{\includegraphics[width=6.0cm,height=5.3cm,keepaspectratio=true]{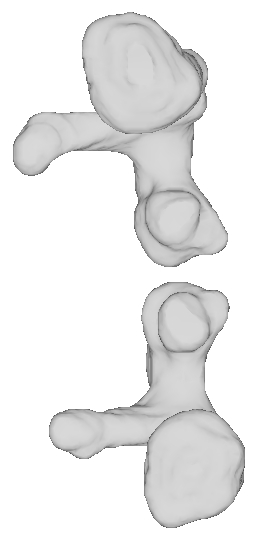}} \par \\[-0.3cm]
\centerline{(a)$\mathcal{M}^0$} \par &
\centerline{(b)$\mathcal{M}^1$} \par &
\centerline{(c)$\mathcal{M}^2$} \par &
\centerline{(d)$\mathcal{M}^3$} \par &
\centerline{(e)$\mathcal{M}_{ref}$} \par \\[-0.5cm]
\end{tabular}
\center
\begin{tabular} { p{0.47\textwidth} p{0.47\textwidth} }
\includegraphics[width=9cm,height=8.0cm,keepaspectratio=true] {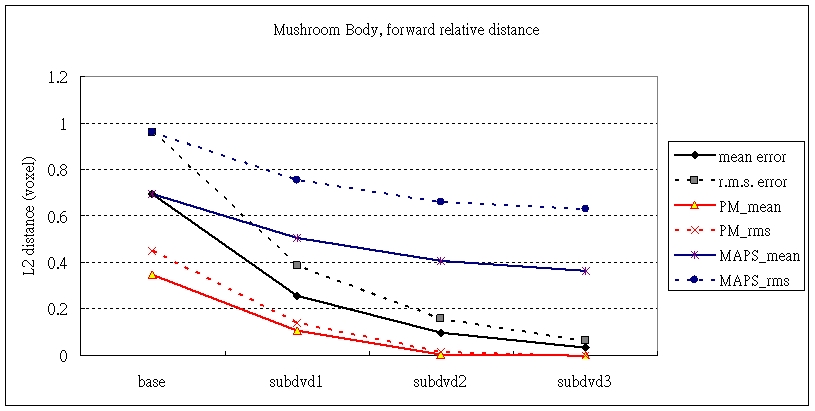}  \par &
\includegraphics[width=9cm,height=4.55cm,keepaspectratio=false]{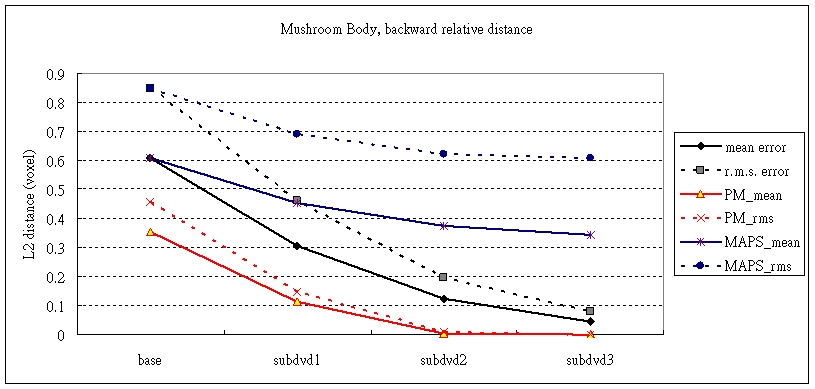}
\par \\[-0.4cm]
 \centerline{(f)} \par &
 \centerline{(g)} \par \\[-0.5cm]
\end{tabular}
\caption{Comparison of errors in meshes derived at different
resolutions. (a)-(d) the coarsest mushroom body mesh
$\mathcal{M}^0$ plus meshes $\mathcal{M}^1$, $\mathcal{M}^2$, and
$\mathcal{M}^3$ derived by our method; (e) the reference mesh; (f)
the forward root-mean-square and mean distances of the reconstructed
meshes compared to those of the reference mesh; (g) the backward
root-mean-square and mean distances of the reference mesh compared
to those of the reconstructed meshes. The \textbf{PM} algorithm yielded the best performance, as shown in (f) and (g). However, for all methods,
the average error per voxel was less than $1$. Since the bounding
box diagonal of the reference mesh was $457$ voxels, less than $1$
voxel difference indicates that the performances of all methods are comparable. 
The number of vertices in the meshes derived by \textbf{PM} is the same as that in the corresponding
$\mathcal{M}^j$. Note that $\mbox{PSNR} = 20 \, \mbox{log}_{10}(\mbox{BBoxDiag}/L^2\mbox{-error})$.
} \label{fig:exp_MB}
\end{figure*}
\begin{figure*}
\center
\begin{tabular}{ p{80pt}p{80pt} p{80pt} p{80pt}p{80pt}}
\centerline{\includegraphics[width=4.61cm,height=4.1cm,keepaspectratio=true]{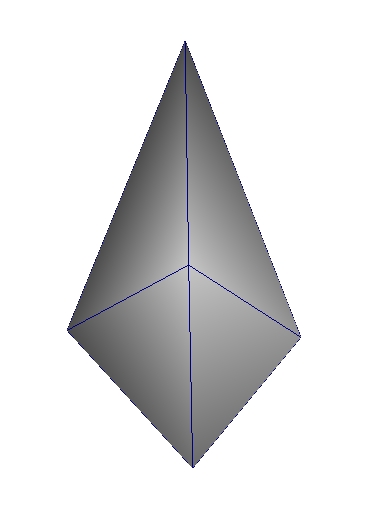}} \par &
\centerline{\includegraphics[width=4.61cm,height=4.1cm,keepaspectratio=true]{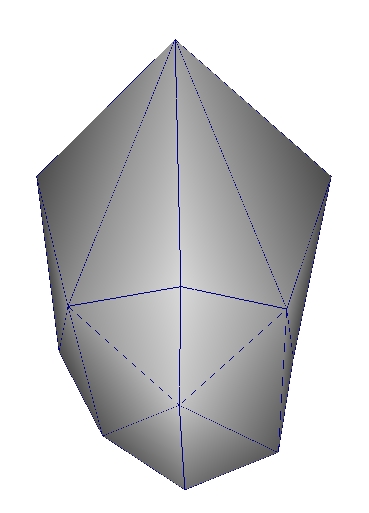}} \par &
\centerline{\includegraphics[width=4.61cm,height=4.1cm,keepaspectratio=true]{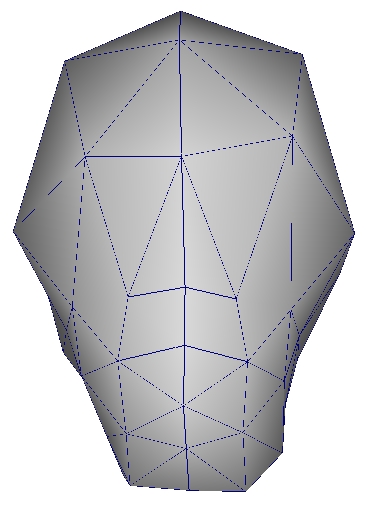}} \par &
\centerline{\includegraphics[width=4.61cm,height=4.1cm,keepaspectratio=true]{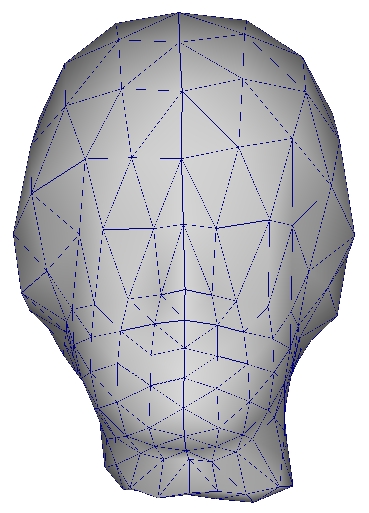}} \par &
\centerline{\includegraphics[width=4.61cm,height=4.1cm,keepaspectratio=true]{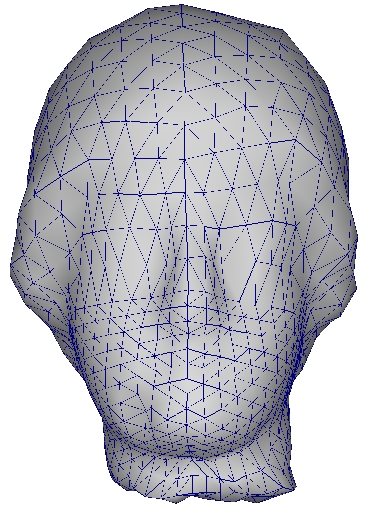}} \par \\[-0.5cm]
\centerline{(a)$\mathcal{M}^0$} \par &
\centerline{(b)$\mathcal{M}^1$} \par &
\centerline{(c)$\mathcal{M}^2$} \par &
\centerline{(d)$\mathcal{M}^3$} \par &
\centerline{(e)$\mathcal{M}^4$} \par \\[-0.5cm]
\centerline{\includegraphics[width=4.61cm,height=4.1cm,keepaspectratio=true]{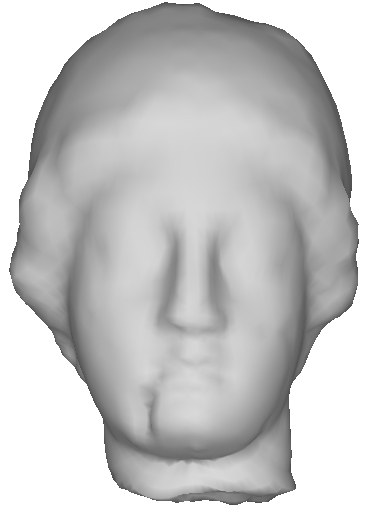}} \par &
\centerline{\includegraphics[width=4.61cm,height=4.1cm,keepaspectratio=true]{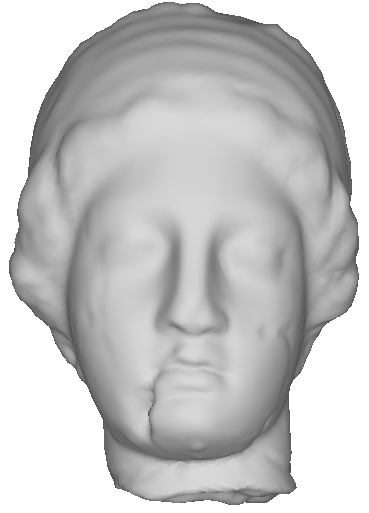}} \par &
\centerline{\includegraphics[width=4.61cm,height=4.1cm,keepaspectratio=true]{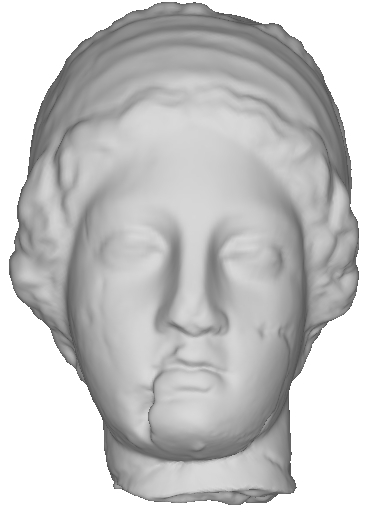}} \par &
\centerline{\includegraphics[width=4.61cm,height=4.1cm,keepaspectratio=true]{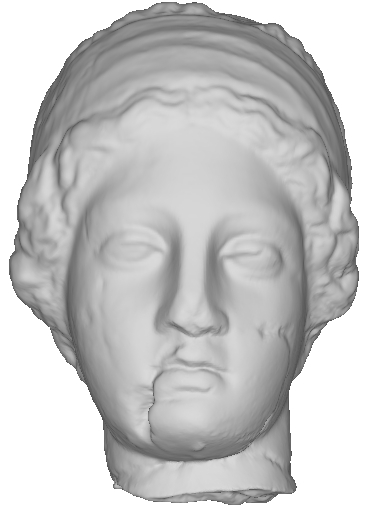}} \par &
\centerline{\includegraphics[width=4.61cm,height=4.1cm,keepaspectratio=true]{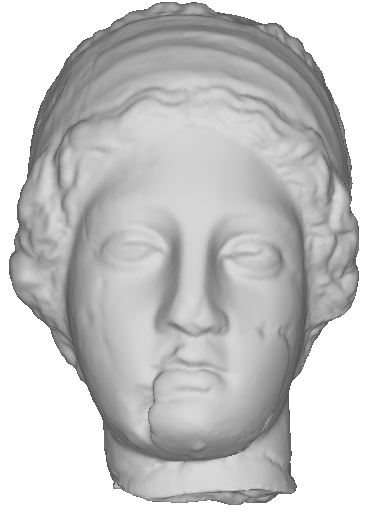}} \par \\[-0.5cm]
\centerline{(f)$\mathcal{M}^5$} \par &
\centerline{(g)$\mathcal{M}^6$} \par &
\centerline{(h)$\mathcal{M}^7$} \par &
\centerline{(i)$\mathcal{M}^8$} \par &
\centerline{(j)$\mathcal{M}_{ref}$} \par \\[-0.6cm]
\end{tabular}
\caption{The reconstruction results of \textbf{Venus Head}. From left to right, top to bottom: the base-domain octahedron $\mathcal{M}^0$, the subdivision results $\mathcal{M}^j \mbox{,} \, j=1 \sim 8$, and the reference surface model $\mathcal{M}_{ref}$. The numbers of vertices and patches of $\mathcal{M}^j$ are $2^{2j+2}+2$ and $8 \times 2^{2j}$ respectively; and the reference contains
$100,759$ vertices and $201,514$ patches.} \label{fig:exp_VE}
\end{figure*}

\begin{figure*}
\center
\begin{tabular}{ p{80pt}p{80pt}p{80pt}p{80pt}p{80pt} }
\centerline{\includegraphics[width=3.61cm,height=3.6cm,keepaspectratio=true]{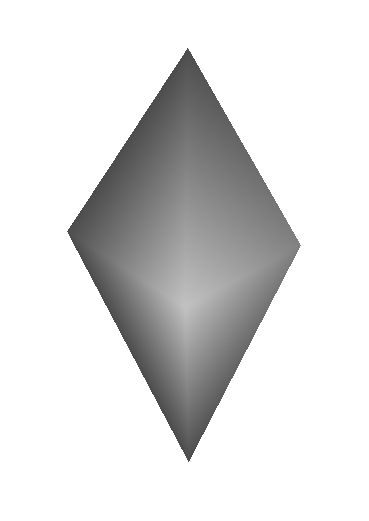}} \par &
\centerline{\includegraphics[width=3.61cm,height=3.6cm,keepaspectratio=true]{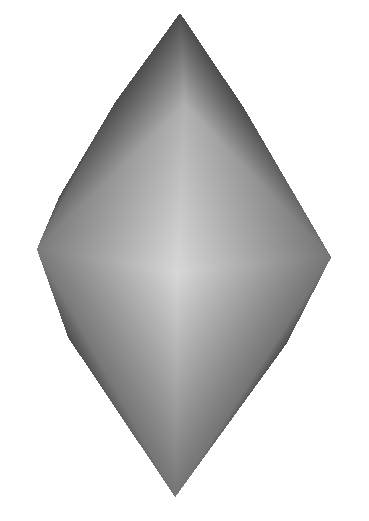}} \par &
\centerline{\includegraphics[width=3.61cm,height=3.6cm,keepaspectratio=true]{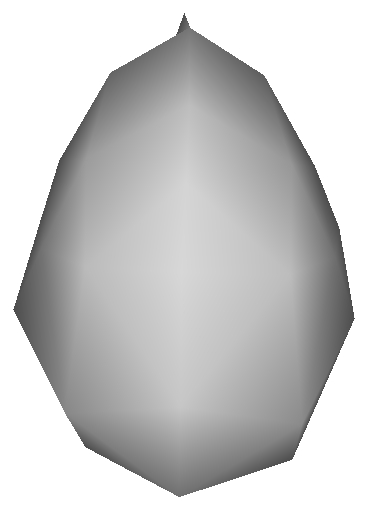}} \par &
\centerline{\includegraphics[width=3.61cm,height=3.6cm,keepaspectratio=true]{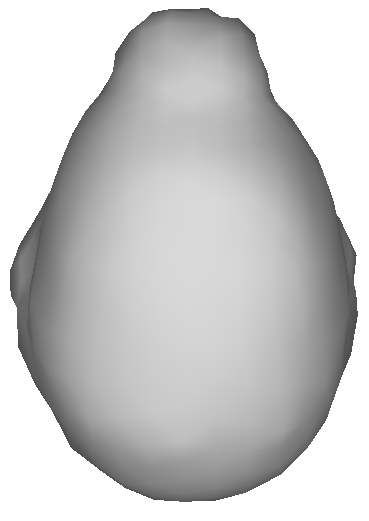}} \par &
\centerline{\includegraphics[width=3.61cm,height=3.6cm,keepaspectratio=true]{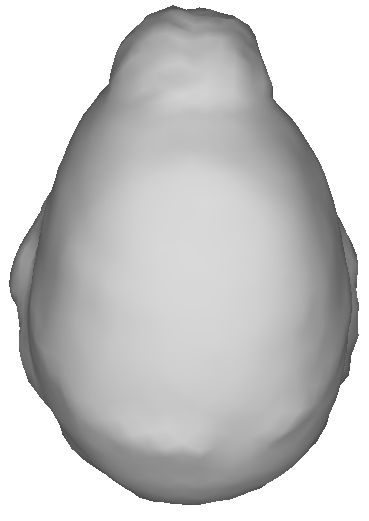}} \par \\ [-0.5cm] 
\centerline{$\mathcal{M}^0_{venus}$} \par &
\centerline{$\mathcal{M}^1_{venus}$} \par &
\centerline{$\mathcal{M}^2_{venus}$} \par &
\centerline{$\mathcal{M}^4_{venus}$} \par &
\centerline{$\mathcal{M}^5_{venus}$} \par \\ [-0.4cm] 
\centerline{\includegraphics[width=3.61cm,height=3.6cm,keepaspectratio=true]{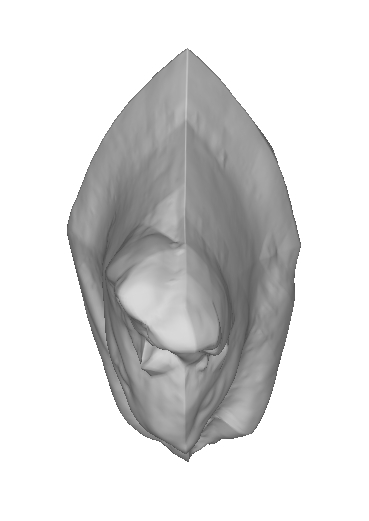}} \par &
\centerline{\includegraphics[width=3.61cm,height=3.6cm,keepaspectratio=true]{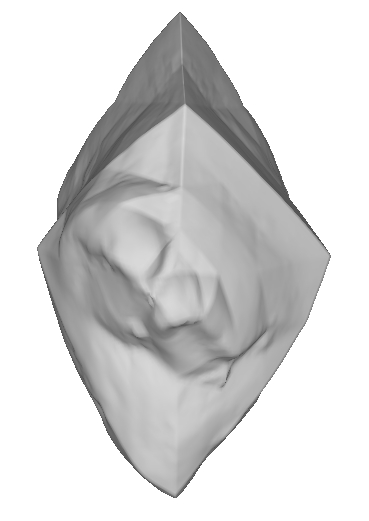}} \par &
\centerline{\includegraphics[width=3.61cm,height=3.6cm,keepaspectratio=true]{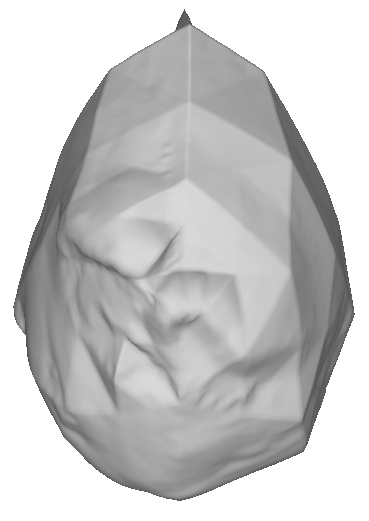}} \par &
\centerline{\includegraphics[width=3.61cm,height=3.6cm,keepaspectratio=true]{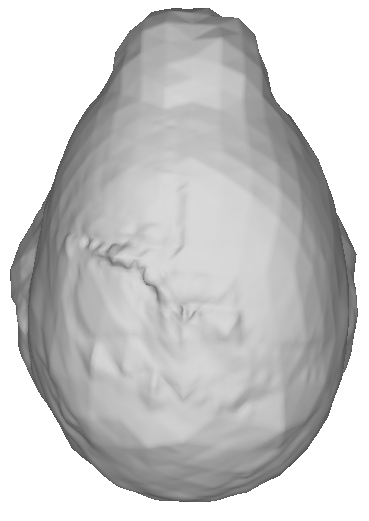}} \par &
\centerline{\includegraphics[width=3.61cm,height=3.6cm,keepaspectratio=true]{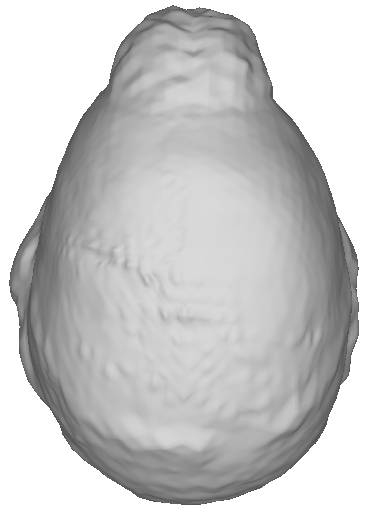}} \par \\ [-0.6cm]
\centerline{\includegraphics[width=3.61cm,height=3.6cm,keepaspectratio=true]{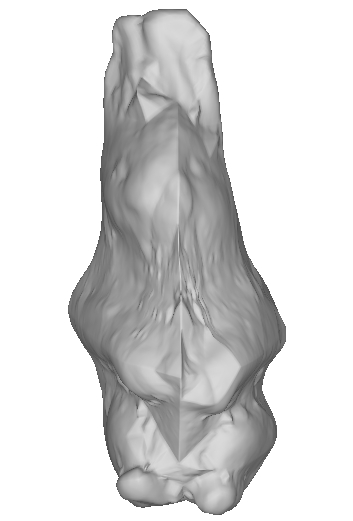}} \par &
\centerline{\includegraphics[width=3.61cm,height=3.6cm,keepaspectratio=true]{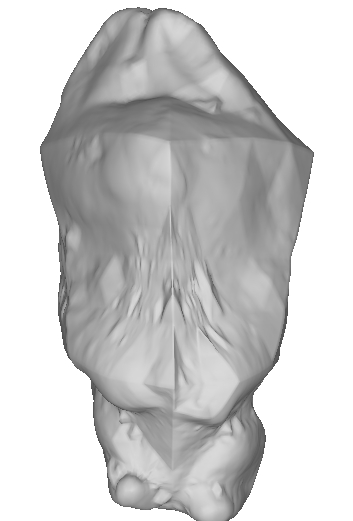}} \par &
\centerline{\includegraphics[width=3.61cm,height=3.6cm,keepaspectratio=true]{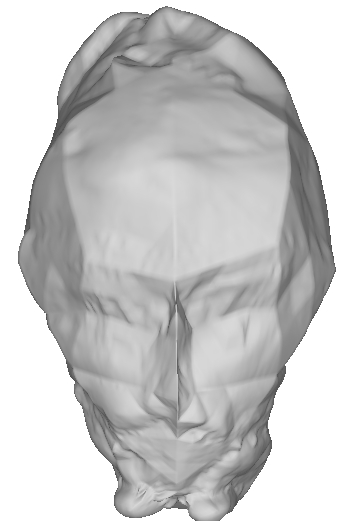}} \par &
\centerline{\includegraphics[width=3.61cm,height=3.6cm,keepaspectratio=true]{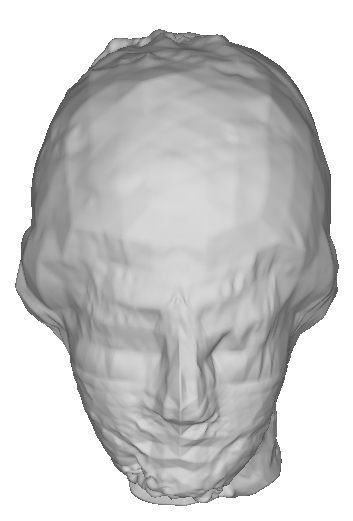}} \par &
\centerline{\includegraphics[width=3.61cm,height=3.6cm,keepaspectratio=true]{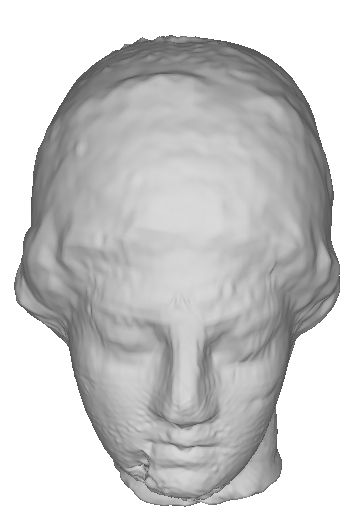}} \par \\[-0.7cm]
\end{tabular}
\caption{Synthesis results of "Vebbit". The coarser meshes are from \textbf{Venus}, and the subdivision coefficients are from \textbf{Rabbit},
that is, $\mathcal{M}^{j-1}_{venus} \bigoplus \mathcal{W}^{j-1}_{rabbit}
\bigoplus \mathcal{W}^{j}_{rabbit} \bigoplus \dots \bigoplus
\mathcal{W}^7_{rabbit}$.
From left to right, $j=1, 2, 3, 5 \mbox{ and } 6$. 
The top row shows the $\mathcal{M}^{j-1}_{venus}$, which serves as the beginning of the reconstruction shown in each column; and the bottom two rows show the synthesis results of different views.}
\label{fig:test_wavelet}
\end{figure*}

\subsection{Remesh and multi-resolution representation} 
\label{application1}

\begin{figure*}[!b]
\begin{tabular}{ p{220pt} p{220pt}}
\includegraphics[width=8cm,height=8cm,keepaspectratio=true]{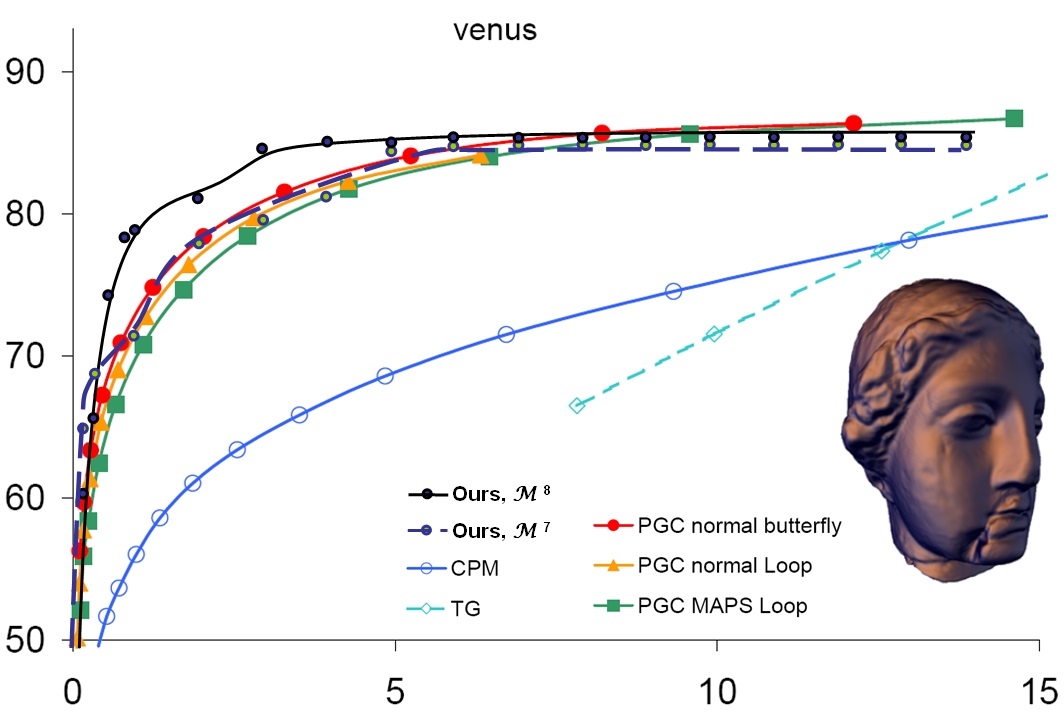} \par &
\includegraphics[width=8cm,height=8cm,keepaspectratio=true]{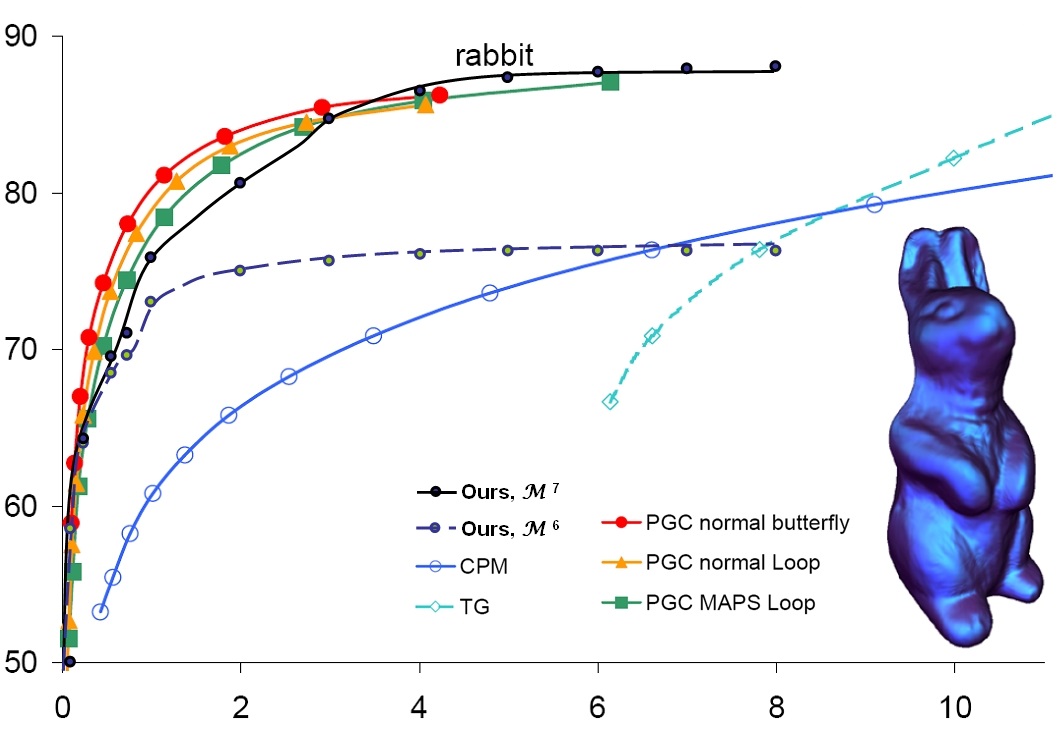} \par \\[-0.5cm]
\centerline{(a)} \par &
\centerline{(b)}  \par \\[-0.4cm]
\includegraphics[width=8cm,height=8cm,keepaspectratio=true]{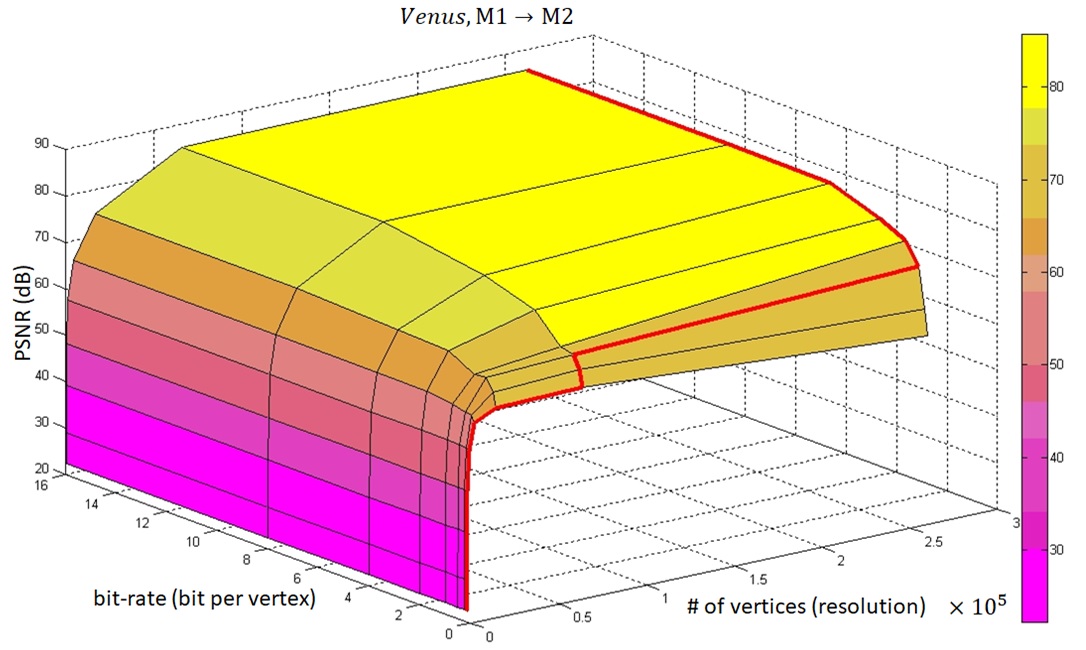} \par &
\includegraphics[width=8cm,height=8cm,keepaspectratio=true]{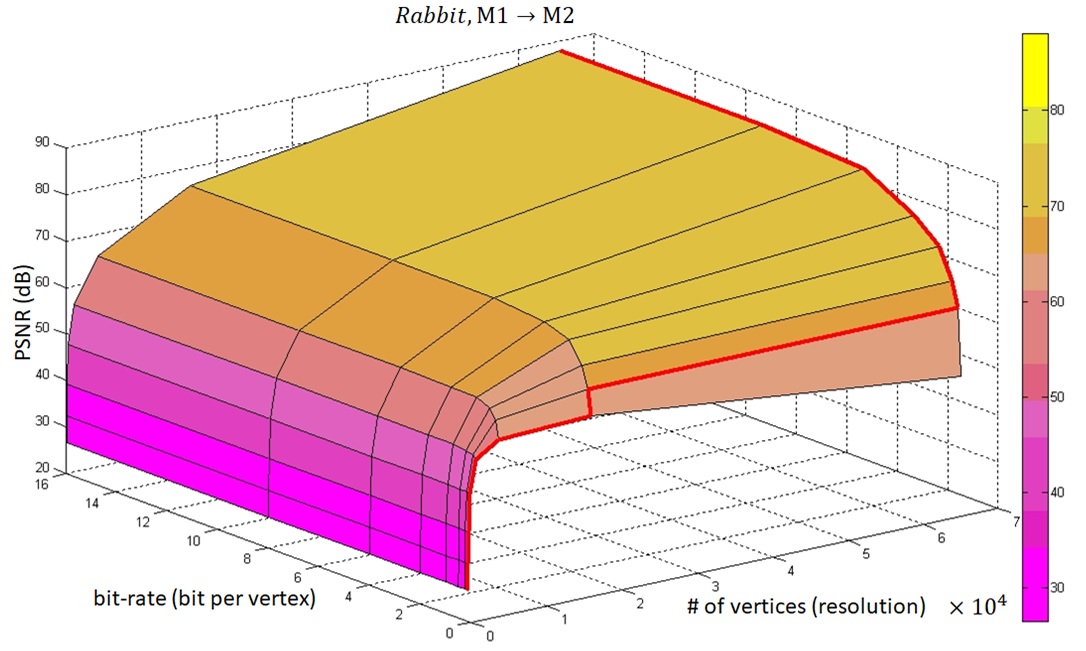} \par \\[-0.5cm]
\centerline{(c)} \par & \centerline{(d)} \par \\[-0.5cm]
\end{tabular}
\begin{tabular}{ p{88pt}p{88pt} p{88pt} p{88pt}p{88pt}}
\centerline{\includegraphics[width=4.61cm,height=3.6cm,keepaspectratio=true]{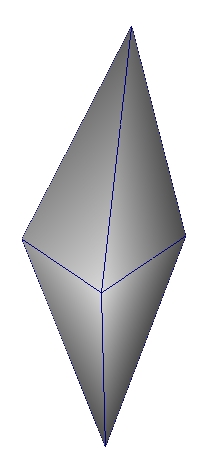}} \par &
\centerline{\includegraphics[width=4.61cm,height=3.6cm,keepaspectratio=true]{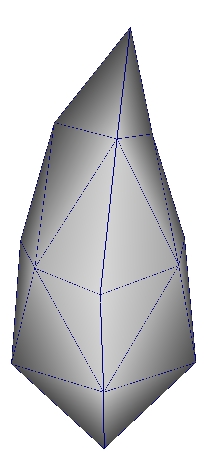}} \par &
\centerline{\includegraphics[width=4.61cm,height=3.6cm,keepaspectratio=true]{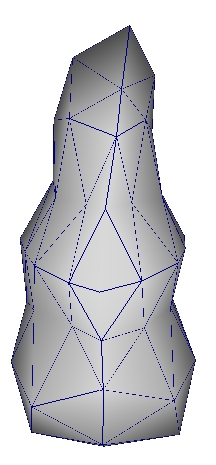}} \par &
\centerline{\includegraphics[width=4.61cm,height=3.6cm,keepaspectratio=true]{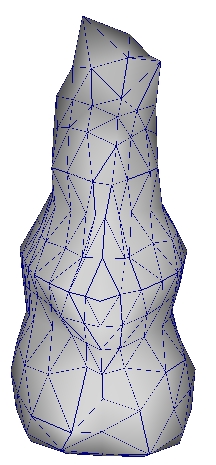}} \par &
\centerline{\includegraphics[width=4.61cm,height=3.6cm,keepaspectratio=true]{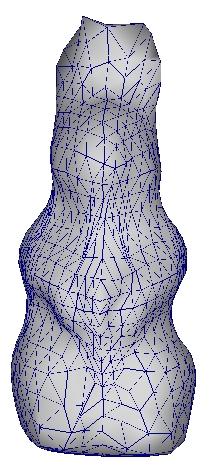}} \par \\[-0.5cm]
\centerline{(e)$\mathcal{M}^0$} \par &
\centerline{(f)$\mathcal{M}^1$} \par &
\centerline{(g)$\mathcal{M}^2$} \par &
\centerline{(h)$\mathcal{M}^3$} \par &
\centerline{(i)$\mathcal{M}^4$} \par \\[-0.4cm]
\centerline{\includegraphics[width=4.61cm,height=3.6cm,keepaspectratio=true]{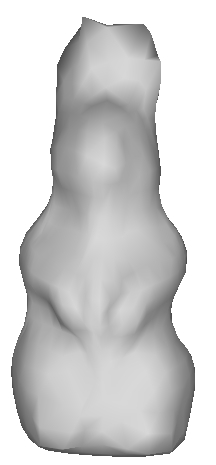}} \par &
\centerline{\includegraphics[width=4.61cm,height=3.6cm,keepaspectratio=true]{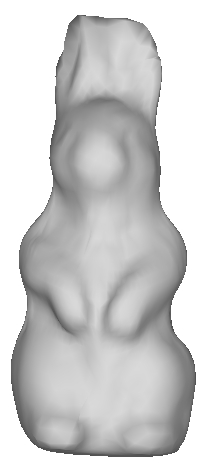}} \par &
\centerline{\includegraphics[width=4.61cm,height=3.6cm,keepaspectratio=true]{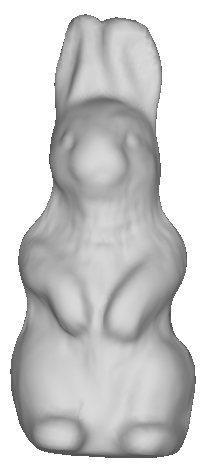}} \par &
\centerline{\includegraphics[width=4.61cm,height=3.6cm,keepaspectratio=true]{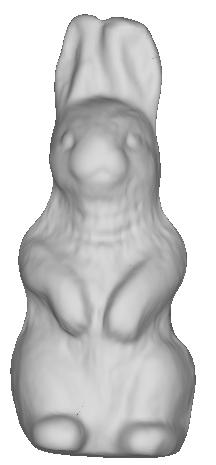}} \par &
\centerline{\includegraphics[width=4.61cm,height=3.6cm,keepaspectratio=true]{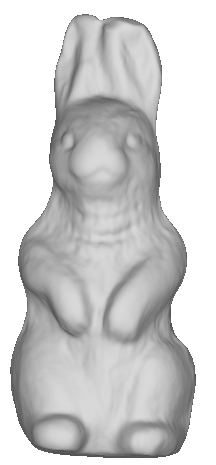}} 
\par \\[-0.4cm]
\centerline{(j)$\mathcal{M}^4$} \par &
\centerline{(k)$\mathcal{M}^5$} \par &
\centerline{(l)$\mathcal{M}^6$} \par &
\centerline{(m)$\mathcal{M}^7$} \par &
\centerline{(n)$\mathcal{M}_{ref}$} \par \\[-0.7cm]
\end{tabular}
\caption{Comparison of the scalable coding performance on \textbf{Venus Head} and \textbf{Rabbit}. 
Sub-figures (a)-(b) are the bpv-PSNR curves of various methods. BWR's performance is comparable to that of the state-of-the-art methods. Sub-figures (c)-(d) show the three-dimensional bpv-PSNR-resolution curves of BWR. The thick red curve denotes the recommended transmission strategy for the model. Sub-figures (e)-(n) show the BWR reconstruction results of \textbf{Rabbit}.
} \label{fig:many_bit_rate_fig}
\end{figure*}

First, we compared the remesh performance of BWR with that of \textbf{PM} and \textbf{MAPS}. The error between the original mesh and the reconstructed mesh was measured by \textbf{METRO}. 
Figures \ref{fig:exp_MB}(a)-(d) show the mesh models of a mushroom body, and a neuropil with a pair of tri-axial structures in the \emph{Drosophila} brain \cite{Shao2014TBME}, reconstructed by our method at different resolutions. 
The original reference mesh shown in Figure \ref{fig:exp_MB}(e) was reconstructed from a $1024 \times 1024 \times 120$ confocal image stack by Chen's method \cite{Chen07}.
The errors measured between the reference mesh and the reconstructed mesh of various mesh representations are compared in Figures \ref{fig:exp_MB}(f) and (g). Our method's performance was between those of \textbf{MAPS} and \textbf{PM}\footnote{\textbf{PM} (\textit{Progressive Mesh}) method is not a remesher. However, since the vertex removal strategy of \textbf{PM} is to remove iteratively a vertex whose absence results in minimal change in current mesh \textbf{geometry}, a simplified mesh derived by \textbf{PM} is regarded to the best approximation of the source mesh at that resolution.}. 
Moreover, Figures \ref{fig:exp_MB}(f) and (g) show that the average error per voxel of the three methods was less than $1$ because the bounding box diagonal for the mushroom body model is $457$ voxels. 
Therefore, the performances of these methods are comparable. 
Finally, note that the $L_2$-errors of the remeshed $\mathcal{M}^8$ \textbf{Venus Head} and $\mathcal{M}^7$ \textbf{Rabbit} models derived by BWR (shown in Figures \ref{fig:exp_VE} and \ref{fig:many_bit_rate_fig}) are $0.002883$ and $0.001435$ with respect to the bounding box diagonal lengths $55.08$ and $36.47$ respectively. The models' PSNR values are thus $85.62$dB and $88.10$dB respectively. The remesh performance of BWR is the same as that of \textbf{NM} and \textbf{MAPS}. 


Next, we consider the semi-regular approximation results of BWR. 
Figure \ref{fig:exp_VE} shows the multi-resolution approximation of the \textbf{Venus Head} model. 
Because BWR can start from a base mesh that has the same genus as the original mesh, we chose an octahedron as the base mesh. The experiment results indicate that different kinds of shape components and details appear in different scales. From $\mathcal{M}^1$ to $\mathcal{M}^3$, the global shape of the input model is reconstructed, whereas facial features and detailed components are reconstructed when $\mathcal{M}^4 \sim \mathcal{M}^6$ and $\mathcal{M}^7 \sim \mathcal{M}^8$ respectively.

\begin{figure*}[b]
\center
\begin{tabular}{ p{80pt}p{80pt}p{80pt}p{80pt}p{80pt} }
\centerline{\includegraphics[width=3.61cm,height=3.6cm,keepaspectratio=true]{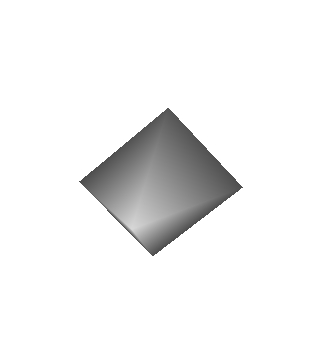}} \par &
\centerline{\includegraphics[width=3.61cm,height=3.6cm,keepaspectratio=true]{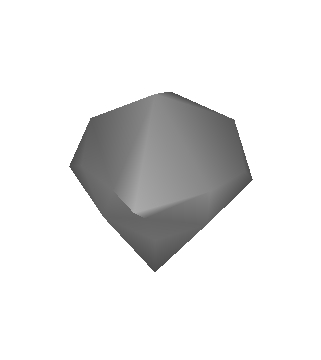}} \par &
\centerline{\includegraphics[width=3.61cm,height=3.6cm,keepaspectratio=true]{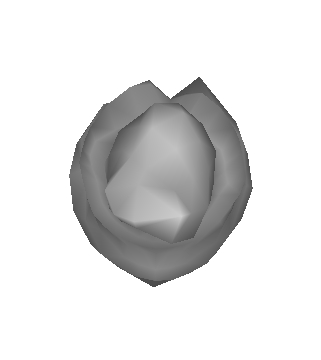}} \par &
\centerline{\includegraphics[width=3.61cm,height=3.6cm,keepaspectratio=true]{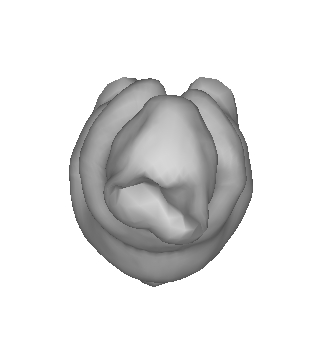}} \par &
\centerline{\includegraphics[width=3.61cm,height=3.6cm,keepaspectratio=true]{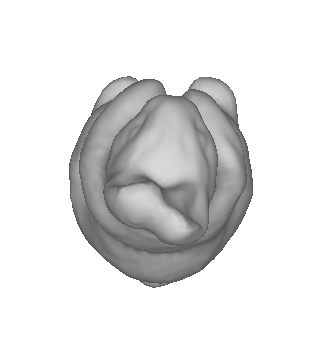}} \par \\[-0.8cm]
\centerline{$\mathcal{M}^0_{rabbit}$} \par &
\centerline{$\mathcal{M}^1_{rabbit}$} \par &
\centerline{$\mathcal{M}^3_{rabbit}$} \par &
\centerline{$\mathcal{M}^5_{rabbit}$} \par &
\centerline{$\mathcal{M}^6_{rabbit}$} \par \\[-0.4cm]
\centerline{\includegraphics[width=3.61cm,height=3.6cm,keepaspectratio=true]{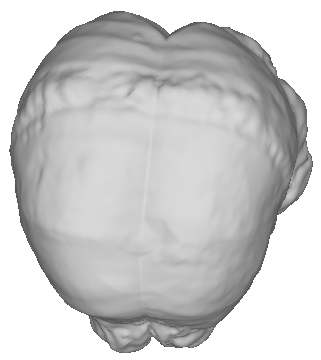}} \par &
\centerline{\includegraphics[width=3.61cm,height=3.6cm,keepaspectratio=true]{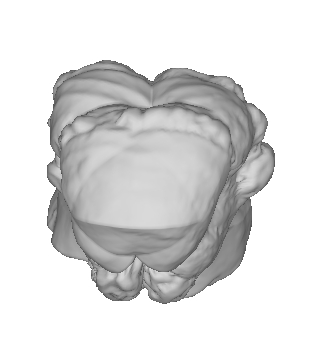}} \par &
\centerline{\includegraphics[width=3.61cm,height=3.6cm,keepaspectratio=true]{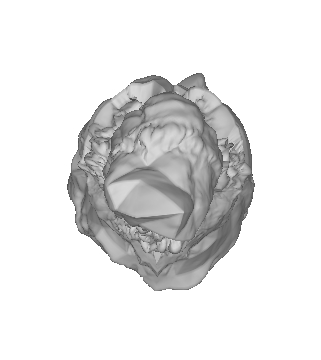}} \par &
\centerline{\includegraphics[width=3.61cm,height=3.6cm,keepaspectratio=true]{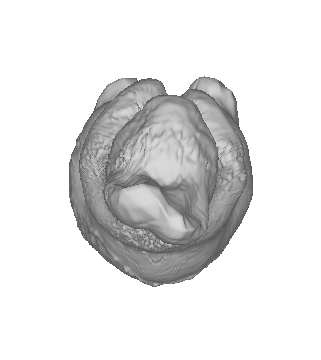}} \par &
\centerline{\includegraphics[width=3.61cm,height=3.6cm,keepaspectratio=true]{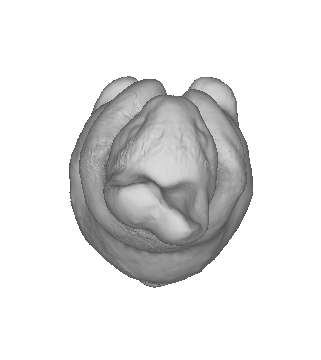}} \par \\[-0.5cm] 
\centerline{\includegraphics[width=3.61cm,height=3.6cm,keepaspectratio=true]{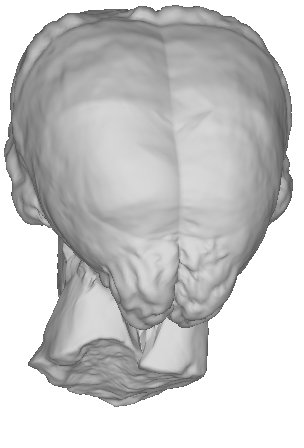}} \par &
\centerline{\includegraphics[width=3.61cm,height=3.6cm,keepaspectratio=true]{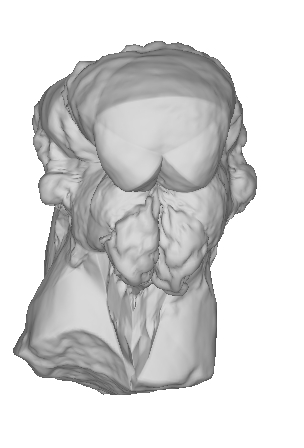}} \par &
\centerline{\includegraphics[width=3.61cm,height=3.6cm,keepaspectratio=true]{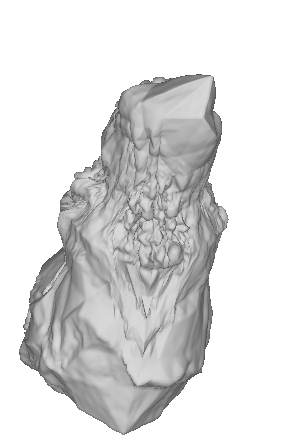}} \par &
\centerline{\includegraphics[width=3.61cm,height=3.6cm,keepaspectratio=true]{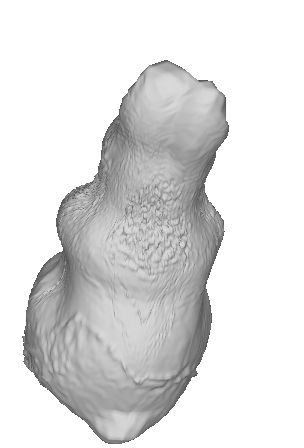}} \par &
\centerline{\includegraphics[width=3.61cm,height=3.6cm,keepaspectratio=true]{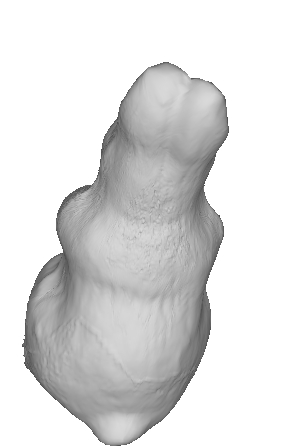}} \par \\[-0.7cm]
\end{tabular}
\caption{Synthesis results of "Ranus". The coarser meshes are from
\textbf{Rabbit}, and the subdivision coefficients are from \textbf{Venus}.
In this example, $j=1, 2, 4, 6 \mbox{ and } 7$ respectively.}
\label{fig:test_wavelet2}
\end{figure*}
\begin{figure*}[t]
\center
\begin{tabular}{ p{130pt} p{130pt} p{130pt}}
\centerline{\includegraphics[width=3.5cm,height=5.8cm,keepaspectratio=true]{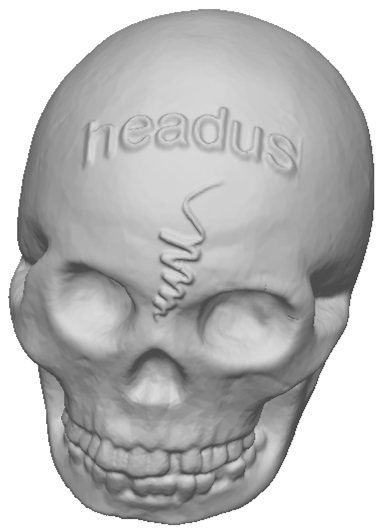}} \par &
\centerline{\includegraphics[width=4.8cm,height=5.8cm,keepaspectratio=true]{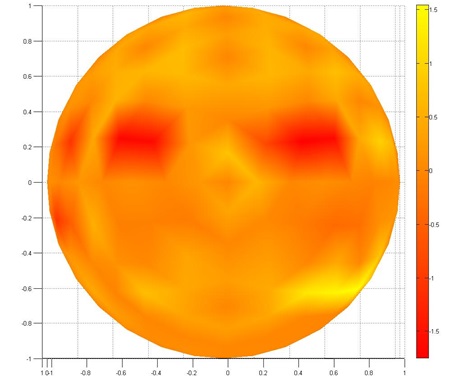}} \par &
\centerline{\includegraphics[width=4.8cm,height=5.8cm,keepaspectratio=true]{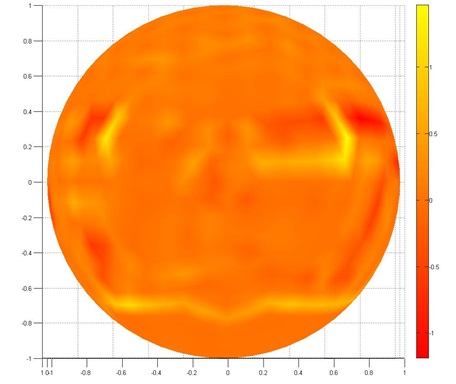}} \par \\[-0.4cm]
\centerline{(a) \textbf{Skull}} \par &
\centerline{(b) $\mathbf{W}^3$ } \par &
\centerline{(c) $\mathbf{W}^4$ } \par \\[-0.4cm]
\centerline{\includegraphics[width=4.8cm,height=5.8cm,keepaspectratio=true]{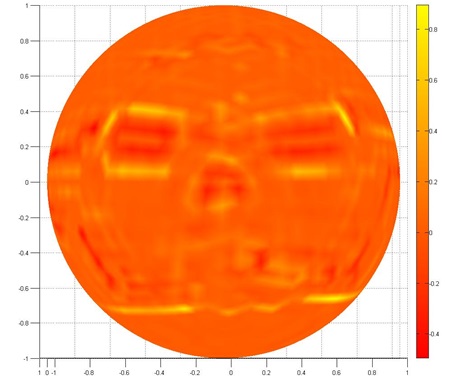}} \par &
\centerline{\includegraphics[width=4.8cm,height=5.8cm,keepaspectratio=true]{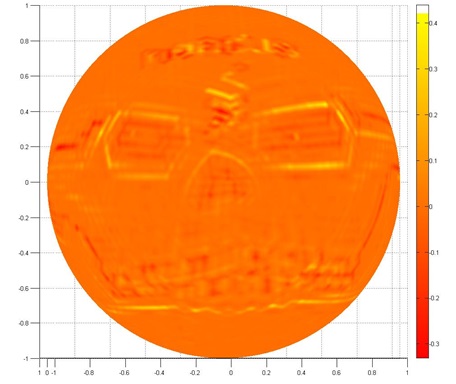}} \par &
\centerline{\includegraphics[width=4.8cm,height=5.8cm,keepaspectratio=true]{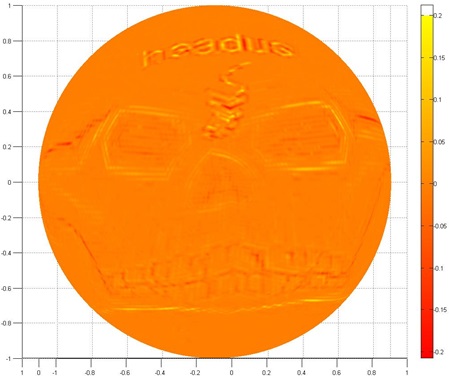}} \par  \\[-0.4cm]
\centerline{(d) $\mathbf{W}^5$ } \par &
\centerline{(e) $\mathbf{W}^6$ } \par &
\centerline{(f) $\mathbf{W}^7$ } \par  \\ [-0.6cm]
\end{tabular}
\caption{Visualization of subdivision coefficients derived from the \textbf{Skull} model. Sub-figures (b)-(f) were generated by MATLAB's built-in function \textbf{trisurf(Tri, X, Y, Z, C)}, where \textbf{(Tri, X, Y, Z)} defines the $j$-th scale unit sphere that the subdivision coefficient matrix $\mathbf{W}^j$ (i.e., the input argument \textbf{C}) was mapped to.
}
 \label{fig:coef_visu}
\end{figure*}

The following experiment results demonstrate that (1) the proposed method can achieve multiresolution representation; and (2) the behavior of $\mathbf{W}^j$ is similar to that of conventional wavelet coefficients.  
Let $\mathbf{W}^j$ denote the subdivision coefficients used to reconstruct $\mathcal{M}^{j+1}$ from $\mathcal{M}^j$. We borrow the ``direct sum'' symbol $\bigoplus$ to link the subdivision procedure and the coefficient matrix.
Then, in Figure \ref{fig:test_wavelet}, each column represents a synthesis result of $\mathcal{M}^{j-1}_{venus} \bigoplus \mathbf{W}^{j-1}_{rabbit} \bigoplus \mathbf{W}^{j}_{rabbit} \bigoplus \dots \bigoplus \mathbf{W}^7_{rabbit}$.
The top row shows the $\mathcal{M}^{j-1}_{venus}$; and the middle and bottom rows show, respectively, the top views and front views of the same synthesis results. Figure \ref{fig:test_wavelet2} illustrates the results of synthesizing $\mathcal{M}^{j-1}_{rabbit}$ and $\mathbf{W}_{venus}^j$, and the bottom row is the rear view.
These experiment results demonstrate that the obtained coefficients can describe how the surface changes from a coarse scale to a fine scale.
Figure \ref{fig:test_wavelet} shows that if we give the subdivision coefficients of \textbf{Rabbit} to \textbf{Venus Head}'s base-domain earlier, the synthesized \textbf{Vebbit} would be more similar to \textbf{Rabbit}.
This finding implies that the subdivision coefficients obtained in coarser resolutions represent global shape components, whereas those obtained in finer resolutions represent local details. Moreover, based on Figure \ref{fig:test_wavelet2}, we find that the hair bud can be roughly decomposed into two components: shape (the left two columns) and textures (the right three columns). The smaller local variations of the texture components make the \textbf{Rabbit} model Godzilla-like from the rear view. 
This observation suggests that the proposed method could be applicable to mesh editing, provided that the user-specified region of interest and the corresponding subdivision coefficients are given.

\begin{figure*}
\begin{tabular}{ p{75pt} p{75pt} p{75pt}p{75pt} p{75pt} p{75pt}}
\centerline{\includegraphics[width=3cm,height=3.5cm,keepaspectratio=true]{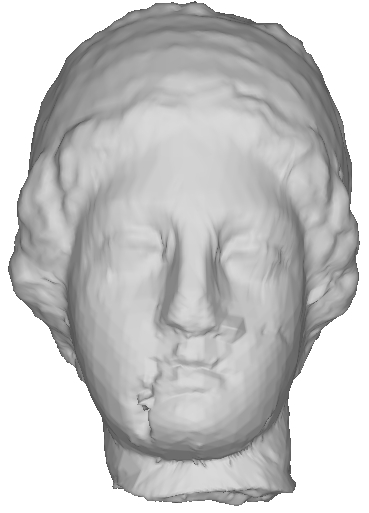}} \par &
\centerline{\includegraphics[width=3cm,height=3.5cm,keepaspectratio=true]{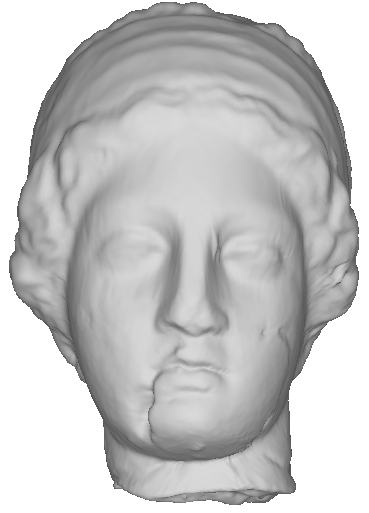}} \par &
\centerline{\includegraphics[width=3cm,height=3.5cm,keepaspectratio=true]{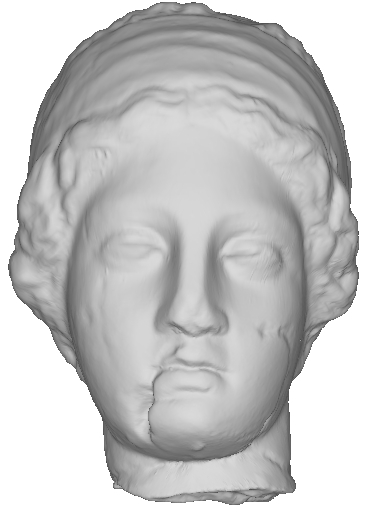}} \par &
\centerline{\includegraphics[width=3cm,height=3.5cm,keepaspectratio=true]{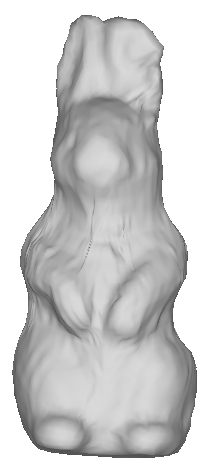}} \par &
\centerline{\includegraphics[width=3cm,height=3.5cm,keepaspectratio=true]{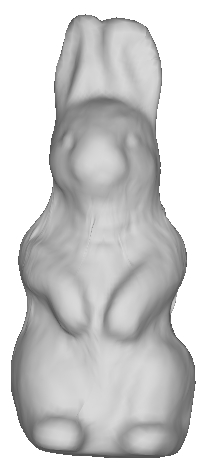}} \par &
\centerline{\includegraphics[width=3cm,height=3.5cm,keepaspectratio=true]{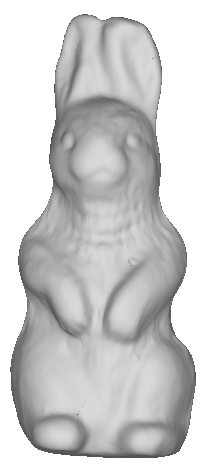}} \par  \\[-0.3cm]
\centerline{(a)$\mathcal{M}^7$, 0.125 bpv} \par &
\centerline{(b)$\mathcal{M}^7$, 0.5 bpv} \par &
\centerline{(c)$\mathcal{M}^8$, 2.0 bpv} \par &
\centerline{(d)$\mathcal{M}^6$, 0.125 bpv} \par &
\centerline{(e)$\mathcal{M}^6$, 0.5 bpv} \par &
\centerline{(f)$\mathcal{M}^7$, 2.0 bpv} \par  \\ [-0.5cm]
\end{tabular}
\caption{Comparison of the visual qualities of the \textbf{Venus Head} and
\textbf{Rabbit} meshes derived by BWR at different bpv values. The meshes
derived at $0.125$ bpv and $0.5$ bpv can be distinguished (the ears
and chest in the \textbf{Rabbit} mesh, and the mouth and surface texture in
the \textbf{Venus Head} mesh); however, meshes derived at $0.5$ bpv and
$2.0$ bpv are almost indistinguishable.} \label{fig:exp_VERA}
\end{figure*}
\begin{figure*}
\center
\begin{tabular} { p{230pt}p{230pt} }
\includegraphics[width=8.5cm,height=11.0cm,keepaspectratio=true]{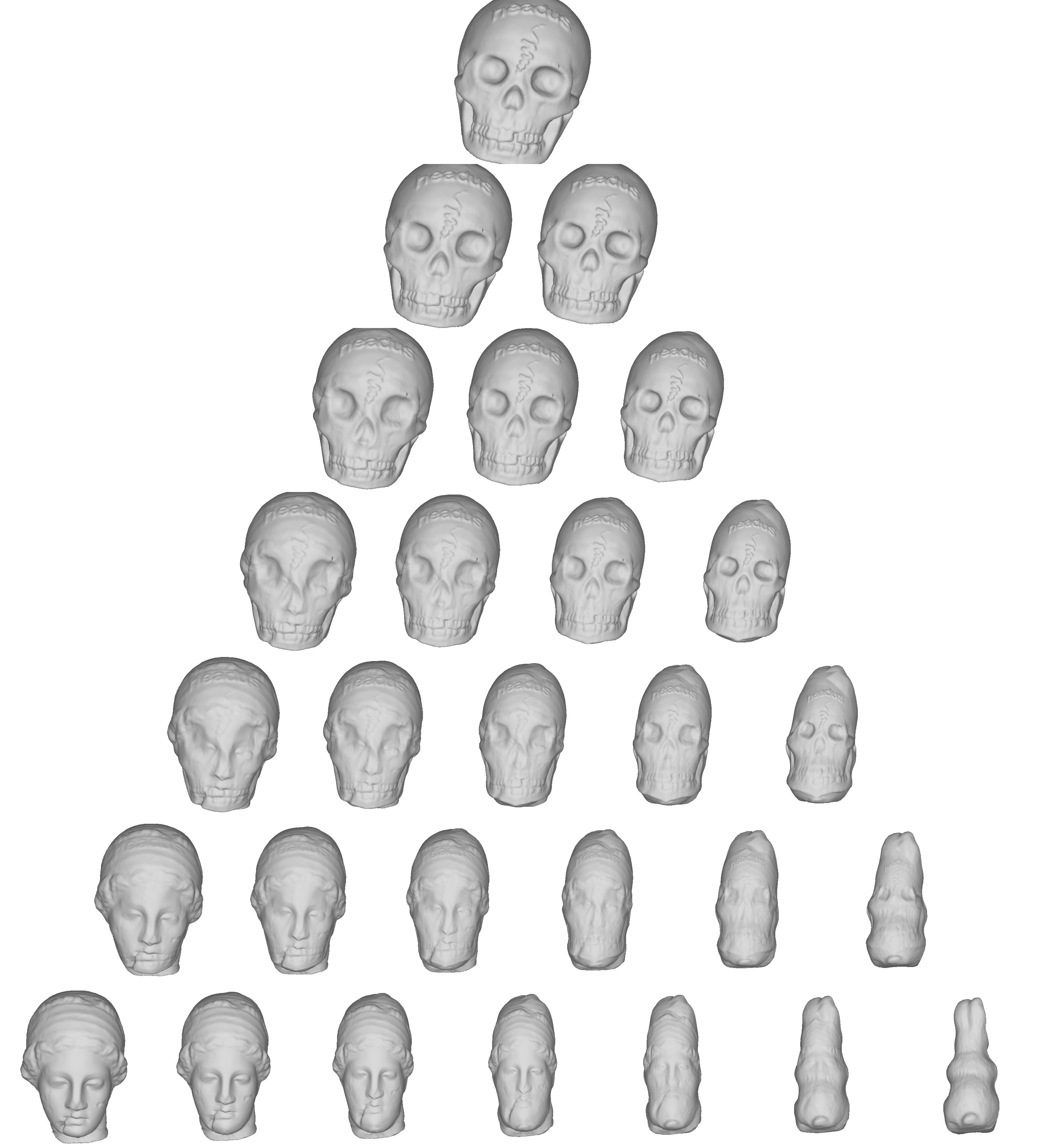} \par &
\includegraphics[width=8.5cm,height=11.0cm,keepaspectratio=true]{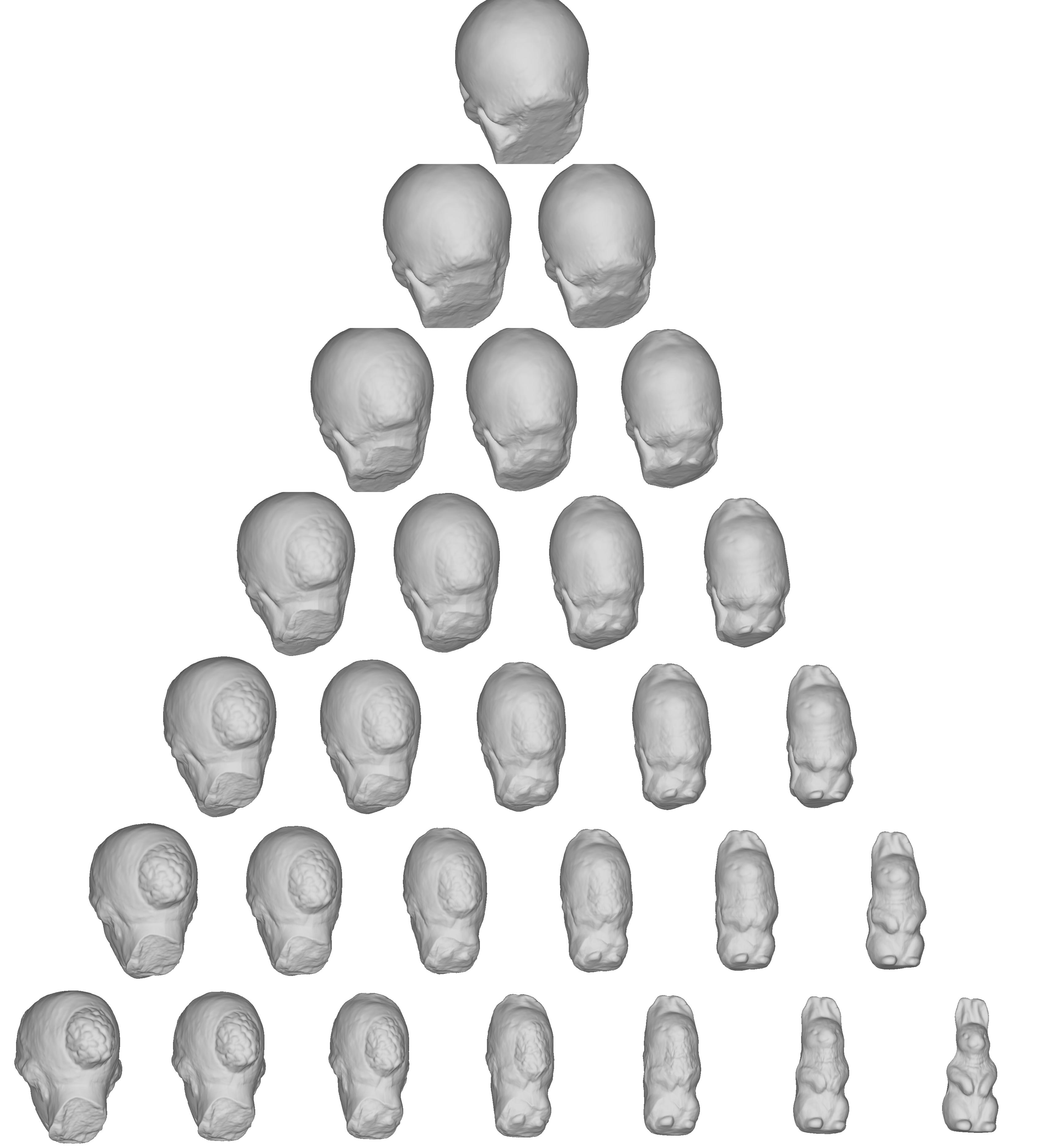} \par \\ [-0.5cm]
\centerline{(a)} \par & \centerline{(b)} \par \\ [-0.7cm]
\end{tabular}
\caption{Morphing sequences of the \textbf{Skull}, \textbf{Rabbit}, and \textbf{Venus Head} models. (a) Front view. (b) Back view.} \label{fig:morphing}
\end{figure*}
\begin{figure}
\centering
\includegraphics[width=0.49\textwidth,height=13.0cm,keepaspectratio=true]{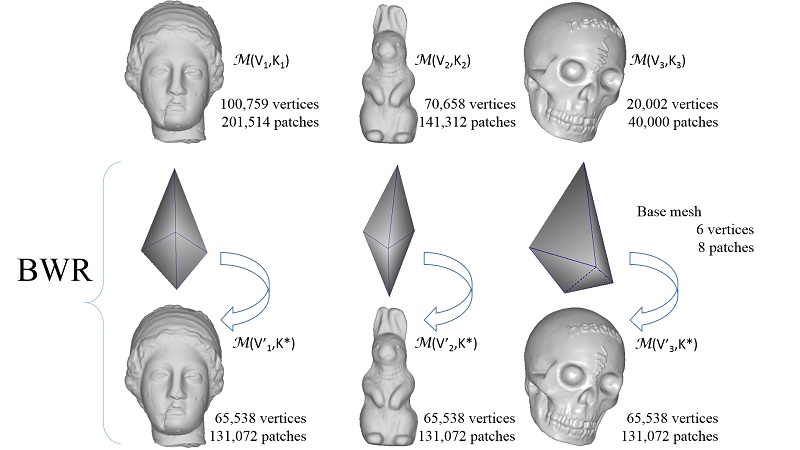}
\caption{The remeshed results derived by BWR share the same topological structure, i.e., $\mathbf{K}^*$. In the figure, $\mathbf{V}_i$ and $\mathbf{K}_i$ are, respectively, the vertices' coordinates and the connectivity information of each mesh model. Consequently, we posit that BWR acts as if it is a transformation procedure capable of converting input meshes into a standard reference space.
} \label{fig:bwr_morph}
\end{figure}
Finally, Figure \ref{fig:coef_visu} shows that BWR can capture local features in a multiscale fashion. 
In contrast to the experiment results presented in Figures {\ref{fig:test_wavelet} and \ref{fig:test_wavelet2}, we mapped each $\mathbf{W}^j$ to a unit sphere and then visualized it to highlight the local features of each scale.
The \textbf{Skull} model in Figure \ref{fig:coef_visu}(a) has five features: 1) eye socket areas, 2) nose area, 3) teeth, 4) jaw, and 5) ``headus" and ``snake-like pattern".
The visualization of $\mathbf{W}^3$ in Figure \ref{fig:coef_visu}(b) proves once again that subdivision coefficients of coarser scales represent global shape structures. The coefficients within the eye socket areas are all negative so that they can make these regions appear concave. 
Thereafter, $\mathbf{W}^4$ sharpens the boundaries of the eye sockets and jaw. 
Positive subdivision coefficients in these zones lift up newly-interpolated vertices to raise the local surface.
Meanwhile, $\mathbf{W}^5$ characterizes the shapes of the teeth, generates the boundaries of the nose, and continues to sharpen the eye sockets and jaw areas. 
The two layers of the coefficients $\mathbf{W}^6$ and $\mathbf{W}^7$ represent very local details, such as the  ``headus" and ``snake-like pattern" on the forehead and boundaries of teeth. We also find that the fringes in Figures  \ref{fig:coef_visu}(e)-(f) are thinner and narrower than those in Figures \ref{fig:coef_visu}(c)-(d) because $\mathbf{W}^6$ and $\mathbf{W}^7$ represent finer structures on the mesh surface.
To summarize, the experiment results presented in this subsection show that BWR can generate a multiscale approximation of the input reference mesh, as well as decompose the input reference mesh to capture local features in a multiscale manner. Such local and multiresolutional features can further enable users to evaluate structural commonness and variations of 3D surface models. Examples of deriving 3D shape properties via BWR can be found in \cite{Shao2017Globalsip}.

\subsection{Scalable Coding} \label{application2}

We compared the coding performance of our method with that of existing methods on two benchmark models:  \textbf{Venus Head} and \textbf{Rabbit}. Our implementation uses the SPIHT algorithm for scalable coding \cite{SPIHT96}. The source code for SPIHT can be downloaded from  the official Matlab forum, Matlab Central \cite{mathworks-spiht}. To ensure a fair comparison, we did not implement the compared methods. Instead, we used the performance curves of those methods provided in \cite{Kho01} and compared them with the curves derived by our method, as shown in Figures \ref{fig:many_bit_rate_fig}(a) and (b). The measurements of our curves were derived from a sequence of mesh approximations ($\mathcal{M}^7$ and $\mathcal{M}^8$ for \textbf{Venus Head}; and, $\mathcal{M}^6$ and $\mathcal{M}^7$ for \textbf{Rabbit}). The multi-resolution approximations of \textbf{Rabbit} from $\mathcal{M}^0$ to $\mathcal{M}^7$ are shown in the third and fourth rows of Figure \ref{fig:many_bit_rate_fig}.

Figures \ref{fig:many_bit_rate_fig}(c) and (d) show, respectively, the three-dimensional (bpv-PSNR-resolution) performance curves for the \textbf{Venus Head} and \textbf{Rabbit} meshes. 
On both curves, the PSNR first increases with increases in the bpv and resolution, and then forms a plateau when the bpv is slightly more than $0.5$ and the resolution is higher than $j=6$. Thus, the \textbf{Venus Head} and \textbf{Rabbit} meshes can be decoded with a satisfactory mesh when the bpv and resolution are set at $0.5$ and $j=6$ respectively. Figure \ref{fig:exp_VERA} shows that the visual quality of meshes reconstructed at $0.5$ bpv is almost indistinguishable from that derived at a higher bpv.

We observe that, at a low bit rate, our \textbf{Venus Head} curve in Figure \ref{fig:many_bit_rate_fig}(a) has a better PSNR than other curves,  
whereas the PSNR curve of our \textbf{Rabbit} model in Figure \ref{fig:many_bit_rate_fig}(b) is not as good as that of the others. This is because we did not impose any shape constraints while constructing the base mesh, so the surface-to-surface error between our $\mathcal{M}^0$, an octahedron, and the original input mesh is surely larger. Hence, it is reasonable that the PSNR curve of \textbf{Rabbit} is not as good as those of other methods in a low bit-rate environment due to an octahedral base mesh.
This weakness can be corrected by compressing a higher resolution mesh. 
As indicated by Figures \ref{fig:many_bit_rate_fig}(c) and (d), 
the $\mathcal{M}^8$ of the \textbf{Venus Head} model yields a better scalable coding performance at a low bit-rate, whereas the performance of $\mathcal{M}^6$ models is worse, even at a high bit-rate. Note that the PSNR values of the two uncompressed models are about 86dB.
This fact implies that when the bandwidth is limited, it is still possible to raise the PSNR effectively by first sending low bpv subdivision coefficients to increase the total number of vertices, and then refining the subdivision coefficients gradually\footnote{For example, in our experiment an $\mathcal{M}^7$ model contains $65,538$ vertices and an $\mathcal{M}^4$ contains $1,026$ vertices. To transmit an $\mathcal{M}^7$ model in 0.25bpv (bit-per-vertex) environment requires about $16,384$ bits. This amount is almost equivalent to that required to transmit an $\mathcal{M}^4$ model in 16.0 bpv.}. 
This transmission strategy is illustrated by the thick red lines in Figures \ref{fig:many_bit_rate_fig}(c) and (d).

\subsection{Morphing}
\label{application3}



Figures \ref{fig:morphing} 
illustrates the metamorphism of three remeshed models. 
The top apex of the triangle is \textbf{Skull}, the bottom-left apex is \textbf{Venus Head}, and the bottom-right apex is \textbf{Rabbit}. 
We used an octahedron as the base mesh for each input mesh. The morphing objects in the triangle were obtained as follows: 
(1) from each input mesh, six vertices were selected in the order of top, right, front, left, back, and bottom; 
(2) the six vertices were used to construct an octahedron as the base mesh for each input mesh; 
(3) the BWR algorithm was applied to the octahedrons to derive the remeshed results; 
(4) we let the mapping function mentioned in Eq.(\ref{eq:eq24}) be $\Pi(i)=i$ because the remeshed results share the same vertex order and the same topological information; 
and, (5) the vertices of the three remeshed models were blended as follows:  
\begin{equation}
\mathcal{M}_{morph} = \alpha \mathcal{M}_{skull} + \beta \mathcal{M}_{venus} + \gamma \mathcal{M}_{rabbit} \mbox{,}
\end{equation}
where $\alpha$, $\beta$, $\gamma$ are barycentric coefficients of the \textbf{Skull-Venus-Rabbit} triangle system shown in Figure \ref{fig:morphing}.

This experiment shows that BWR can convert input meshes into a standard reference domain, as illustrated in Figure \ref{fig:bwr_morph}. By ordering the apexes, edges, and faces of the base mesh correctly, the user-specified pre-alignment process ensures that the connectivity information of the multiresolution approximation meshes is the same. Therefore, the mesh morphing problem as well as mesh registration and warping can be solved in a straightforward manner.
 
\subsection{Time Complexity} 
\label{subsec:timecomplexity}
The piercing procedure used to determine the subdivision coefficient matrix $\mathbf{W}^j$ causes a computational bottleneck in BWR. 
In practice, the procedure can be computed rapidly if its implementation is optimized. For example, Guskov et al. reported that the \textbf{NM} method, comprised of a base mesh generating phase and a piercing procedure phase, can generate remeshing results containing $59,319$ vertices from a $234$-vertex base mesh in $6.8$ minutes by using a Pentium III CPU \cite{Guskov00}. 
However, because we adopted the most straightforward strategy, i.e., a full-search strategy, to implement the piercing procedure, the computational complexity of our program is theoretically $\mathcal{O}(mf)$, which is higher than that of Guskov et al.'s implementation. Here, ``f" is the number of triangular faces of the input reference mesh, and ``m'' is the number of newly-interpolated vertices of the current scale. Table \ref{tb:timecost} and Figure \ref{fig:timecost} show the execution times of our experiments, each of which begins by reading the $\mathcal{M}^j$ and $\mathcal{M}_{ref}$ files (.txt) and ends by writing the $\mathcal{M}^{j+1}$ file (.txt) to the hard disk. 
The experiment results show that the computational complexity of our full-search-based implementation is $\mathcal{O}(mf)$, an upper bound for the piercing procedure. 

In sum, the computational time cost of BWR can be reduced to several minutes if the implementation is optimized in the same way as the \textbf{NM} method. In addition, we should emphasize that BWR can produce multiscale mesh representation and remeshing results simultaneously, whereas the total time cost of the conventional mesh wavelet transform should include the computation times of remeshing and mesh wavelet analysis. In this paper, we do not consider the problem of how to design a data structure to facilitate the piercing procedure. We will address the problem in a future study.

\begin{table}
\center
\begin{tabular}{|rr|r|r|r|}
\multicolumn{5}{l}{\textbf{TABLE} \ref{tb:timecost}} \\
\hline
\multicolumn{2}{|c|}{ Scale and }&  \multicolumn{1}{|c|}{\bf Venus Head} & \multicolumn{1}{|c|}{\bf Rabbit} & \multicolumn{1}{|c|}{\bf Skull}  \\
\multicolumn{2}{|c|}{ \# of new vertices} &  &   &   \\ \hline
$\mathcal{M}^0 \rightarrow \mathcal{M}^1$: & 12& 7.26 & 4.94 & 1.78 \\
$\mathcal{M}^1 \rightarrow \mathcal{M}^2$: & 48& 11.82 & 7.78  & 2.77 \\
$\mathcal{M}^2 \rightarrow \mathcal{M}^3$: & 192& 31.07 &  20.58 & 6.81 \\
$\mathcal{M}^3 \rightarrow \mathcal{M}^4$: & 768& 114.5 & 71.11 & 22.88 \\
$\mathcal{M}^4 \rightarrow \mathcal{M}^5$: & 3,072& 420.5 & 275.3 & 89.92 \\
$\mathcal{M}^5 \rightarrow \mathcal{M}^6$: & 12,288& 1649 & 1093 & 373.3 \\
$\mathcal{M}^6 \rightarrow \mathcal{M}^7$: & 49,152& 6804 & 4478 & 1670 \\
$\mathcal{M}^7 \rightarrow \mathcal{M}^8$: & 196,608& 29689 & 22079 & 9673 \\
\hline
\end{tabular}
\caption{Time costs of the examples discussed in Section \ref{sec:exp}. All times are in seconds on a virtual Windows XP machine with 2GB RAM and one core shared by an Intel i5-3470 CPU. The numbers of vertices and faces of the input references \textbf{Venus Head}, \textbf{Rabbit}, and \textbf{Skull} models are $(100759, 201514)$, $(70658,141312)$, and $(20002,40000)$ respectively.}
\label{tb:timecost}
\end{table}
\begin{figure}
\includegraphics[width=8cm,height=11cm,keepaspectratio=true]{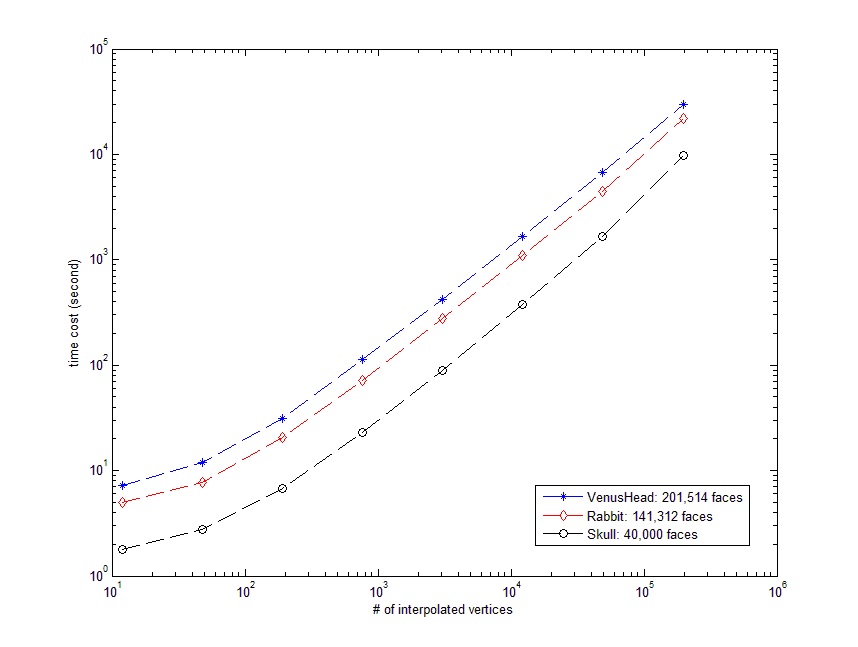}
\caption{Computational time cost. Each data point in the figure includes the program execution time and I/O time.} \label{fig:timecost}
\end{figure}

\section{Limitation and Future Work}
\label{sec:limit}
The limitation of the proposed method is that the $\vec{s}$ and the original input mesh must have a correct intersection. 
For models with conspicuously protruding structures, e.g., the legs of the \textit{horse} model and the ears of the  \textit{bunny} model, it is difficult to generate the multiresolution approximation from a very simple base mesh, e.g., an octahedron. 
There are two possible ways to resolve this issue. The first is to derive a multiresolution representation by BWR from a more complicated base mesh, e.g. a simplified mesh derived by \textbf{PM} or \textbf{MAPS}, similar to the experiment results we described in Figure \ref{fig:exp_MB}. 
The second way is to develop a more robust algorithm to derive and iteratively adjust $\vec{s}$ for each interpolated vertex.
The constraints on developing a robust algorithm for $\vec{s}$
of the $j^{th}$ scale are as follows. 
(1) The $\vec{s}$ of adjacent newly interpolated vertices should vary gradually and homogeneously; that is, the angle between the $\vec{s}$ of any two adjacent vertices should be small so that triangular patches will not be folded. 
(2) The $\vec{s}$ must be derived solely by $\mathcal{M}^j$'s local information; otherwise, the decoder of the scalable transmission will impose an extra overhead, e.g., either an overhead on $\vec{s}$ or the $\mathbf{d}^j$ in Eq.(\ref{perturbation1}), to reconstruct $\mathcal{M}^j$.
The first constraint is the reason that we chose the vertex normal to replace the 
$\vec{s}$ derived by the butterfly scheme in flat areas and regions covering shape edges. The second constraint is optional because such overhead information only influences the efficiency of scalable coding and will not alter the topological structure of our multiresolution mesh.

Given the above limitation and the discussion in Section \ref{subsec:timecomplexity}, there are two aspects of BWR that could be improved in our future work. The first is to develop an algorithm that can guarantee the direction vectors of adjacent newly interpolated vertices will vary gradually and homogeneously. The second is to design a data structure to integrate BWR and the piercing procedure so that the time complexity can be reduced.


\section{Concluding Remarks} 
\label{conclusion}
We have proposed a backward coarse-to-fine method (called BWR) that can construct a multi-resolution approximation of a mesh surface. With the help of the original mesh, our method starts by using any base mesh of the same genus type as the original mesh and terminates in a semi-regular approximation of the original mesh. In addition, we have shown that the locations of new vertices in BWR can be derived as a simple closed form and represented as a scalar. Our experiment results show that this representation is effective for scalable coding and its performance is comparable to or better than that of some state-of-the-art methods. 
We have also shown that BWR can (1) generate a tiling-invariant multiscale mesh representation; and (2) act as if it is a transformation procedure capable of converting the original input mesh into a standard reference domain.
As a result, morphing becomes a straightforward interpolation of corresponding vertices in the approximating meshes derived by BWR. We believe that the simplicity and flexibility of BWR makes it suitable for mesh editing operations and applicable to biomedical applications; for example, tracking deformations of a 3D beating heart model and deriving a 3D calibrated reference space for image atlasing.

%
%
%
%
%
%


\ifCLASSOPTIONcaptionsoff
  \newpage
\fi



\bibliographystyle{plain}

%

%

\begin{IEEEbiography}
[{\includegraphics[width=1.5in,height=1.5in,clip,keepaspectratio,bb=0 0 525 700]{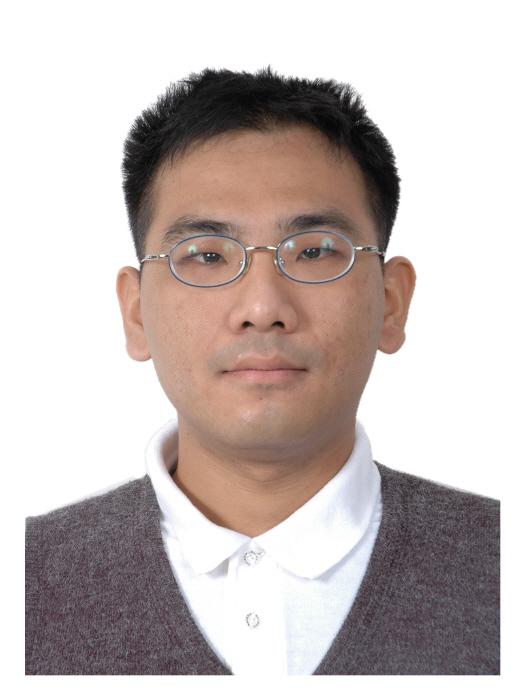}}]
{Hao-Chiang~Shao}
(S’09---M’13) received the B.S., M.S. and Ph.D. degree in electrical engineering in 2001, 2003 and 2012 respectively, from the National Tsing Hua University, Hsinchu, Taiwan, R.O.C. He was a postdoctoral researcher in Inst. Information Science, Academia Sinica in 2012-2017, and joined in a series of \textit{Drosophila} brain research projects.  In 2017-2018, he was an R\&D engineer, in the Computational Intelligence Technology Center, Industrial Technology Research Institute, taking charges of DNN-based automated optical inspection (AOI) projects. 
His research interests include image processing, registration and warping, mesh processing, wavelet analysis, and light field super-resolution. He is currently an assistant professor in the Department of Statistics and Information Science, Fu Jen Catholic University, Taiwan.
\end{IEEEbiography}
\end{document}